\documentclass[%
reprint,
 amsmath,amssymb,
 aps,
 pre,
]{revtex4-2}

\usepackage{graphicx, epstopdf}
\usepackage{dcolumn}
\usepackage{bm}
\usepackage{amssymb}
\usepackage{float}
\usepackage{color}
\usepackage{hyperref}
\usepackage{subfigure}
\usepackage{braket}
\usepackage{cases}
\usepackage{comment}
\usepackage{ulem}

\begin{document}


\title{Characterizing the Hyperuniformity of Disordered Network Metamaterials}


\author{Charles Emmett Maher}
 \email[Email: ]{cemaher@email.unc.edu}
 \affiliation{Department of Mathematics, University of North Carolina at Chapel Hill, Chapel Hill, NC, USA}

\author{Katherine A. Newhall}%
\email[Email: ]{knewhall@email.unc.edu}
\affiliation{Department of Mathematics, University of North Carolina at Chapel Hill, Chapel Hill, North Carolina, USA}

\date{\today}

\begin{abstract}
Advancements in materials design and manufacturing have allowed for the production of ordered and disordered metamaterials with diverse properties.
Hyperuniform two-phase heterogeneous materials, which anomalously suppress density fluctuations on large length scales compared to typical disordered systems, and network materials are two classes of metamaterials that have desirable physical properties.
Recent focus has been placed on the design of disordered hyperuniform network metamaterials that inherit the desirable properties of both of these metamaterial classes.
In this work, we focus on determining the extent to which network structures derived from the spatial tessellations of hyperuniform point patterns inherit the hyperuniformity of the progenitor point patterns.
In particular, we examine the Delaunay, Voronoi, Delaunay-centroidal, and Gabriel tessellations of nonhyperuniform and hyperuniform point patterns in two- and three-dimensional Euclidean space.
We use the spectral density to characterize the density fluctuations of two-phase media created by thickening the edges of these tessellations in two dimensions and introduce a variance-based measurement to characterize the network structures directly in two and three dimensions.
We find that, while none of the tessellations completely inherit the hyperuniformity of the progenitor point pattern, the degree to which the hyperuniformity is inherited is sensitive to the tessellation scheme and the short- and long-range translational disorder in the point pattern, but not to the choice of beam shape when mapping the networks into two-phase media.

\end{abstract}

\maketitle


\section{Introduction}\label{sec:Intro}
Advancements in materials design and manufacturing have allowed for the production of ordered and disordered metamaterials with diverse properties including, e.g., negative refractive indices \cite{smith_metamaterials_2004}, negative thermal expansion \cite{dubey_negative_2024}, or the ability to mimic the mechanical properties of biological tissues \cite{sniechowski_heterogeneous_2015, mohammed_design_2018, ramirez-torres_three_2018, wit_simulation_2019}.
Disordered hyperuniform materials are a particularly interesting class of metamaterial, in which infinite-length scale density fluctuations are anomalously suppressed compared to those in ordinary disordered systems \cite{zachary_hyperuniformity_2009,torquato_hyperuniform_2018}.
Such materials have been demonstrated to have many desirable physical properties, such as complete isotropic photonic band gaps \cite{florescu_designer_2009, froufe-perez_role_2016, gkantzounis_hyperuniform_2017, siedentop_stealthy_2024} as well as optimal transport \cite{torquato_diffusion_2021, torquato_multifunctional_2018} and mechanical properties \cite{torquato_multifunctional_2018,kim_new_2019}.
A second class of exciting metamaterials is network materials, whose low density and better strength-to-weight and stiffness-to-weight ratios \cite{mcgregor_manufacturability_2024} than bulk solids make them ideal for use in, e.g., medical and aerospace applications.
Due to the advantages of these material classes, there have been recent efforts to design and fabricate disordered hyperuniform network metamaterials to capture the beneficial properties from both of these material classes \cite{sniechowski_heterogeneous_2015, wit_simulation_2019, obrero_electrical_2024, siedentop_stealthy_2024}.

Previous work has focused extensively on the characterization of the structure and properties of hyperuniform point patterns and two-phase heterogeneous materials in $d$-dimensional Euclidean space $\mathbb{R}^d$ (see, e.g., Ref. \citenum{torquato_hyperuniform_2018} and references therein).
In particular, a hyperuniform point pattern is one in which the local number density of particles $\sigma_N^2(R)$ associated with a spherical observation window of radius $R$ grows more slowly than $R^d$ \cite{torquato_local_2003}.
Equivalently, a point pattern is hyperuniform if the structure factor $S(\mathbf{k})$ tends to zero as the wavenumber $k\equiv|\mathbf{k}|$ tends to zero.
Likewise, a two-phase material is hyperuniform if its local volume-fraction variance $\sigma_V^2(R)$ decays more quickly than $R^{-d}$ or if its spectral density $\tilde{\chi}_{_V}(\mathbf{k})$ tends to zero as the wavenumber $k$ tends to zero \cite{zachary_hyperuniformity_2009}.
Of particular interest are disordered \textit{stealthy} hyperuniform systems \cite{torquato_ensemble_2015}, whose relevant spectral functions are identically zero for some set of wavenumbers around the origin.
To assess the degree of hyperuniformity of computer- or laboratory-generated structures, one can use the hyperuniformity index $H$ \cite{chen_binary_2018}, defined as the value of the spectral density $\tilde{\chi}_{_V}(k)$ extrapolated to $k=0$ divided by its global maximum, where a system is \textit{effectively} hyperuniform if $H$ is below some empirical threshold (often $10^{-2}$ \cite{chen_binary_2018, maher_characterization_2022}).

Due to the great multitude of possible hyperuniform two-phase structures, there is interest in devising fast and simple methods for generating readily manufacturable hyperuniform structures. 
One possible avenue for doing so is the direct mapping of a hyperuniform point pattern to a two-phase material.
While it is known that, e.g., the mapping of a hyperuniform point pattern to a packing of identical nonoverlapping spheres preserves the hyperuniformity of the progenitor point pattern \cite{torquato_disordered_2016}, it is not generally known what mappings do or do not preserve hyperuniformity.
Due to the desirable photonic \cite{man_isotropic_2013,siedentop_stealthy_2024}, elastic, and conductive \cite{torquato_multifunctional_2018} properties of network structures derived from hyperuniform point patterns, there is a reasonable expectation that network structures inherit the hyperuniformity of their progenitor point patterns.
Indeed, in Ref. \citenum{torquato_multifunctional_2018}, Torquato and Chen suggest that such structures may be hyperuniform due to the presence of these desirable physical properties, but note that the exact conditions required to generate truly hyperuniform networks from disordered hyperuniform point patterns are yet to be identified.

There has been a recent focus on characterizing the structures of networks generated via modifying existing point patterns in various ways.
In Ref. \cite{torquato_multifunctional_2018}, Torquato and Chen showed network materials derived from tessellations of disordered stealthy hyperuniform point patterns have near-optimal elastic and conductive physical properties.
In addition, they demonstrated that different tessellation schemes are better for certain physical properties than others, however, the large-scale structural properties of these network materials were not directly assessed.
Similarly, Raj \textit{et al.} \cite{raj_upcoming_2025} examine the transport properties of networks derived from the tessellations of hyperuniform and nonhyperuniform point patterns, as well as their local geometric properties.
Moreover, in Refs. \citenum{chen_stonewales_2021} and \citenum{chen_topological_2021}, the authors show that disordered network structures generated by adding topological defects to ordered two-dimensional networks have hyperuniform vertex locations and unique electronic transport properties.
Salvalaglio \textit{et al.} \cite{salvalaglio_persistent_2024a} demonstrated that, by analyzing the topology of networks generated via persistent homology, one can diagnose the hyperuniformity of finite point patterns with a particular class of hyperuniform and nearly hyperuniform structure factors.
Their method, however, does not examine the hyperuniformity of the networks generated with these methods.
Moreover, Newby \textit{et al.} have characterized the local fluctuations in Voronoi cell volumes \cite{newby_structural_2024} and density fluctuations in the nodes of Delaunay networks \cite{newby_point_2025} derived from hyperuniform and nonhyperuniform point patterns using several correlation functions and network characterization methods.
We also note here that Kim and Torquato \cite{kim_new_2019} have devised a method to produce large disordered hyperuniform particle dispersions based on spatial tessellations of point patterns, but their procedure is distinct from what we consider herein.

These works lack a direct characterization of the large-scale density fluctuations of the structures themselves derived from hyperuniform point patterns.  In this work, for the first time, we characterize the degree to which the hyperuniformity of the point pattern is inherited by network structures composed of the edges of tiles from spatial tessellations of 2D and 3D hyperuniform point patterns (see Sec. \ref{sec:Background} for more details).
Specifically, we build on the configuration-to-network-to-3D printed structure pipeline introduced in Obrero \textit{et al.} \cite{obrero_electrical_2024} for designing and manufacturing disordered network metamaterials.
First, we generate uniformly randomized lattice (URL) \cite{klatt_cloaking_2020} and disordered stealthy hyperuniform point patterns with varying degrees of short-scale translational disorder, as well as totally uncorrelated point patterns in $d = 2,3$ under periodic boundary conditions.
Then, we generate the Voronoi (V), Delaunay (D), Delaunay-Centroidal (C), and Gabriel (G) tessellations of these point patterns and use the edges of the tiles in these tessellations to form spatially embedded network structures (see also \cite{raj_upcoming_2025}).

Subsequently, we characterize these structures in two ways.
First, to characterize the structures of two-phase media derived from 2D networks across length scales, we apply a ``thickness'' to the network edges such that the network fills a non-zero fraction of the space, then compute the spectral density $\tilde{\chi}_{_V}(\mathbf{k})$.
Second, inspired by traditional variance-based order characterization methods \cite{torquato_local_2022, skolnick_quantifying_2024, maher_local_2024}, we propose the use of a variance measure, the edge-length variance $\sigma^2_{\ell}(R)$, associated with the total edge length of a spatially embedded network within a spherical observation window of radius $R$, which we apply to our 2D and 3D networks.
Such a variance measure allows us to analyze the structure of the networks themselves as opposed to two-phase structures derived from them, avoiding the need to make a particular choice for the edge thickness or shape when analyzing the structures.

We find that the hyperuniformity of point patterns is not completely inherited by network structures derived from their spatial tessellations, and the degree to which hyperuniformity is inherited varies across different point patterns, tessellation types, and space dimensions.
Such subtleties in the inheritance of hyperuniformity by the network structures are consistent with the varying rank-orderings of physical properties for the 2D networks derived from different tessellations of hyperuniform and nonhyperuniform point patterns in Ref. \citenum{torquato_multifunctional_2018}.
More specifically, for 2D networks derived from disordered hyperuniform point patterns with small degrees of local translational order, we find that G networks do not inherit the hyperuniformity of the progenitor point patterns, while D, V, and C are only effectively hyperuniform with C networks inheriting hyperuniformity most effectively.  
Furthermore, increasing the local translational disorder does not change the hyperuniformity of the underlying point pattern, but it does increase the values of the hyperuniformity index $H$ for the D, V, and C networks.  For large degrees of local translational disorder, we find all the networks are nonhyperuniform.


We also demonstrate that our proposed measure $\sigma^2_{\ell}(R)$ has analogous scaling behaviors to the number variance $\sigma_N^2(R)$, and detects the same structural characteristics at intermediate- and large-length scales in our network structures as the well-established $\tilde{\chi}_{_V}(k)$ calculation.
This indicates that the incomplete inheritance of hyperuniformity by the two-phase network media is intrinsic to the tessellation-based network structures and not a result of a particular choice of edge thickness (or shape, for $d = 3$) when mapping the network structures to two-phase media.
Having established the correspondence between traditional variance measures and $\sigma_{\ell}^2(R)$ for the 2D networks, we apply the same method to 3D networks and show that V networks derived from disordered hyperuniform point patterns tend to have the greatest suppression of large-scale density fluctuations, followed by C, D, and G networks.
This change in optimal network type from C networks for $d=2$ to V networks for $d=3$ is not unexpected given that the currently known structures that minimize large-scale density fluctuations also change across space dimensions \cite{torquato_local_2003}.

The rest of the paper is organized as follows. Section \ref{sec:Background} contains pertinent mathematical background for the methods used to generate and characterize the point patterns and networks used in this work. In Sec. \ref{Sec:NetworkHU}, we discuss how hyperuniformity can be extended to spatial networks and introduce the edge-length variance $\sigma_{\ell}^2(R)$. In Sec. \ref{sec:Results}, we characterize the structures of these networks using the spectral density and edge-length variance. In Sec. \ref{sec:Conc} we offer concluding remarks and outlook for future studies.

\section{Mathematical Background and Methods}\label{sec:Background}

In this section, we review the quantification of the pair statistics of point patterns and two-phase heterogeneous media and how they can be used to determine and characterize the degree of hyperuniformity in a given system.
Then, we introduce the one nonhyperuniform and two hyperuniform point pattern models considered herein.
Finally, we describe the four tessellation schemes we use to convert these point patterns into networks and the method used to convert these networks into two-phase media.

\subsection{Pair statistics}
A system of point particles in $\mathbb{R}^d$ is completely statistically characterized by the $n$-particle probability density functions $\rho_n(\mathbf{r}_1,\dots,\mathbf{r}_n)$ for all $n\geq1$, which are proportional to the probability of finding $n$ particles at the positions $\mathbf{r}_1,\dots,\mathbf{r}_n$ \cite{hansen_theory_1990}.
For statistically homogeneous systems, $\rho_1(\mathbf{r_1})$ is equal to the number density $\rho$ and $\rho_2(\mathbf{r_1},\mathbf{r_2})=\rho^2g_2(\mathbf{r})$, where $\mathbf{r}=\mathbf{r_2}-\mathbf{r_1}$ and $g_2(\mathbf{r})$ is the pair correlation function.
In cases where the system is also statistically isotropic, $\mathbf{r}$ can be replaced by the scalar radial distance $r$.
The structure factor $S(\mathbf{k})$ is defined as
\begin{equation}
    S(\mathbf{k})=1+\rho\tilde{h}(\mathbf{k}),
\end{equation}
where $\tilde{h}(\mathbf{k})$ is the Fourier transform of the total correlation function $h(\mathbf{r})=g_2(\mathbf{r})-1$ and $\mathbf{k}$ is a wave vector.

One can also characterize the pair statistics in systems of point particles by computing the local number variance $\sigma_N^2(r)$ associated with a spherical observation window of radius $R$.
Specifically, $\sigma_N^2(R)$ can be obtained by directly sampling the number of points $N(R)$ in uniformly randomly positioned observation windows, i.e., $\sigma_N^2(R)\equiv\langle N^2(R)\rangle-\langle N(R)\rangle^2$, or given in terms of $g_2(\mathbf{r})$ or $S(\mathbf{k})$ \cite{torquato_local_2003}:
\begin{equation}\label{eq:nv}
    \begin{split}
        \sigma_N^2 &=\rho v_1(R)\left[1+\rho\int_{\mathbb{R}^d} h(\mathbf{r})\alpha_2(r;R)d\mathbf{r}\right], \\
        &=\rho v_1(R)\left[\frac{1}{(2\pi)^d}\int_{\mathbb{R}^d}S(\mathbf{k})\tilde{\alpha}_2(k;R)d\mathbf{k} \right],
    \end{split}
\end{equation}
where $v_1(R)$ is the volume of a sphere with radius $R$, $\alpha_2(r;R)$ is the intersection volume of two spheres of radius $R$ whose centroids are separated by a distance $r$ scaled by the volume of one such window, and $\tilde{\alpha}_2(k;R)$ is its Fourier transform.

Two-phase heterogeneous media are domains of space $\mathcal{V}\subseteq\mathbb{R}^d$ partitioned into two disjoint regions $\mathcal{V}_1, \mathcal{V}_2$ with volume fractions $\phi_1, \phi_2$ \cite{torquato_random_2002a}.
The microstructures of such media can be completely characterized by the $n$-point probability functions $S_n^{(i)}(\mathbf{r}_1,\dots,\mathbf{r}_n)$ that describe the probability that the points $(\mathbf{r}_1,\dots,\mathbf{r}_n)$ fall in phase $i$ and are defined by \cite{torquato_random_2002a}
\begin{equation}
    S_n^{(i)}(\mathbf{r}_1,\dots,\mathbf{r}_n) = \left\langle\prod_{j=1}^{n}\mathcal{I}^{(i)}(\mathbf{r_j})\right\rangle,
\end{equation}
where $\mathcal{I}^{(i)}(\mathbf{r_j})$ is the indicator function for phase $i$:
\begin{equation}
    \mathcal{I}^{(i)}(\mathbf{x}) =
    \begin{cases}
    1, & \mathbf{x} \in \mathcal{V}_i\\
    0, & \textrm{else}.
    \end{cases}
\end{equation}

For statistically homogeneous media, $S^{(i)}_1(\mathbf{r}_1) = \phi_i$ and the two-point correlation function $S_2^{(i)}(\mathbf{r})$ depends only on the displacement vector $\mathbf{r}$.
The corresponding two-point autocovariance function is obtained by subtracting the long-range behavior from $S_2^{(i)}(\mathbf{r})$:
\begin{equation}
    \chi_{_V}(\mathbf{r}) = S_2^{(1)}(\mathbf{r})-\phi_1^2 = S_{(2)}^2(\mathbf{r})-\phi_2^2.
\end{equation}
The nonnegative spectral density $\tilde{\chi}_{_V}(\mathbf{k})$ is defined as the Fourier transform of $\chi_{_V}(\mathbf{r})$.
One can also compute $\tilde{\chi}_{_V}(\mathbf{k})$ from pixelized or voxelized representations of two-phase media via \cite{zachary_hyperuniformity_2011}
\begin{equation}\label{eq:SpecDen_Pix}
    \tilde{\chi}_{_V}(\mathbf{k})=\frac{|\sum_{j=1}^{N_p}e^{-i\mathbf{k}\cdot{\mathbf{r}_j}}\tilde{m}(\mathbf{k};A)|^2}{V},\;(\mathbf{k}\neq0),
\end{equation}
where $N_p$ is the number of filled pixels (voxels), $\mathbf{r}_j$ is the position of pixel (voxel) $j$, $V$ is the volume of the system, and $\tilde{m}(\mathbf{k};A)$ is the Fourier transform of a pixel (voxel) with geometric parameters $A$.

Like $\sigma_N^2(R)$ for point patterns, there is an analogous variance-based characterization of two-phase media, in particular, the local volume-fraction variance $\sigma_V^2(R)$ associated with a spherical observation window of radius $R$ \cite{zachary_hyperuniformity_2009}.
Again, like $\sigma_N^2(R)$ one can obtain $\sigma_V^2(R)$ via direct sampling or compute it given $\chi_{_V}(\mathbf{r})$ or $\tilde{\chi}_{_V}(\mathbf{k})$:
\begin{equation}\label{Bivolumefractionreal}
\begin{split}
    \sigma_{V}^2(R) &= \frac{1}{v_1(R)} \int_{\mathbb{R}^d}  \chi_{_V}(\textbf{r})\alpha_2(r;R)d\textbf{r},\\
    &= \frac{1}{v_1(R)(2\pi)^d} \int_{\mathbb{R}^d}  
\tilde{\chi}_{_{V}}(\textbf{k}) \tilde{\alpha}_2(k;R)d\textbf{k}.
\end{split}
\end{equation}

\subsection{Hyperuniformity}
Consider a point pattern whose structure factor $S(\mathbf{k})$ scales like a power law in the vicinity of the origin, i.e. $S(\mathbf{k})\sim k^{\alpha}$, where $\alpha$ is the hyperuniformity scaling exponent.
Such a point pattern is hyperuniform if $\alpha > 0$.
Equivalently, a point pattern is hyperuniform if its number variance $\sigma_N^2(R)$ grows more slowly than the observation window volume, i.e.,
\begin{equation}
    \lim_{R\rightarrow\infty}\frac{\sigma_N^2(R)}{v_1(R)} = 0. 
\end{equation}
The hyperuniformity scaling exponent $\alpha$ can be used to divide hyperuniform systems into three distinct classes that describe their associated large-$R$ $\sigma_N^2(R)$ scaling  \cite{torquato_local_2003}:
\begin{equation}\label{eq:BiN_classes}
    \sigma_N^2(R)\sim
    \begin{cases}
        R^{d-1},\quad\alpha > 1 &\textrm{(class I)}\\
        R^{d-1}\textrm{ln}(R),\quad\alpha = 1&\textrm{(class II)}\\
        R^{d-\alpha},\quad0<\alpha<1&\textrm{(class III)},
    \end{cases}
\end{equation}
where class I and III are the strongest and weakest types of hyperuniformity, respectively.
Also contained in class I are the so-called stealthy hyperuniform systems, whose structure factors $S(k)$ are identically 0 for some finite range of wavenumbers around the origin and come in ordered (i.e., crystalline) and disordered varieties. 

Analogously, a two-phase medium is considered hyperuniform if its spectral density $\tilde{\chi}_{_V}(k)$ tends to 0 as the wavenumber tends to 0 or, equivalently, if $\sigma_V^2(R)$ decays more rapidly than the reciprocal of the window-volume growth rate, i.e.,
\begin{equation}\lim_{R\rightarrow\infty}\sigma_V^2(R)v_1(R)=0.
\end{equation}
Like with point patterns, one can sort hyperuniform two-phase media into three different classes based on the hyperuniformity scaling exponent $\alpha$ for media whose spectral densities behave like a power law in the vicinity of the origin \cite{torquato_disordered_2016}:
\begin{numcases}{\sigma_{_V}^2(R) \sim }\label{BiV_classesI}
  R^{-(d+1)}, \quad \alpha >1  & \(  (\text{class I})\) \nonumber \\
  R^{-(d+1)}\ln(R),\quad \alpha = 1 & \( (\text{class II})\)   \\
  R^{-(d+\alpha)}, \quad (0 < \alpha < 1) & \( (\text{class III}).\)   \nonumber
\end{numcases}
In practice, a two-phase medium is considered to be effectively hyperuniform if \cite{chen_binary_2018, maher_characterization_2022}
\begin{equation}
    H = \frac{\tilde{\chi}_{_V}(0)}{\tilde{\chi}_{_V}(k_p)} \leq 10^{-2},
\end{equation}
where $\tilde{\chi}_{_V}(0)$ is the value of the spectral density when extrapolated to the origin and $\tilde{\chi}_{_V}(k_p)$ is the value of the largest peak of the spectral density.
Here, measurements of $H$ are done using angular-averaged $\tilde{\chi}_{_V}(k)$ with a bin size of $0.5k_{min}$, where $k_{min}$ is the smallest accessible wavenumber.

\begin{figure*}[!t]
    \centering
        \subfigure[]{\includegraphics[height=0.4\textwidth]{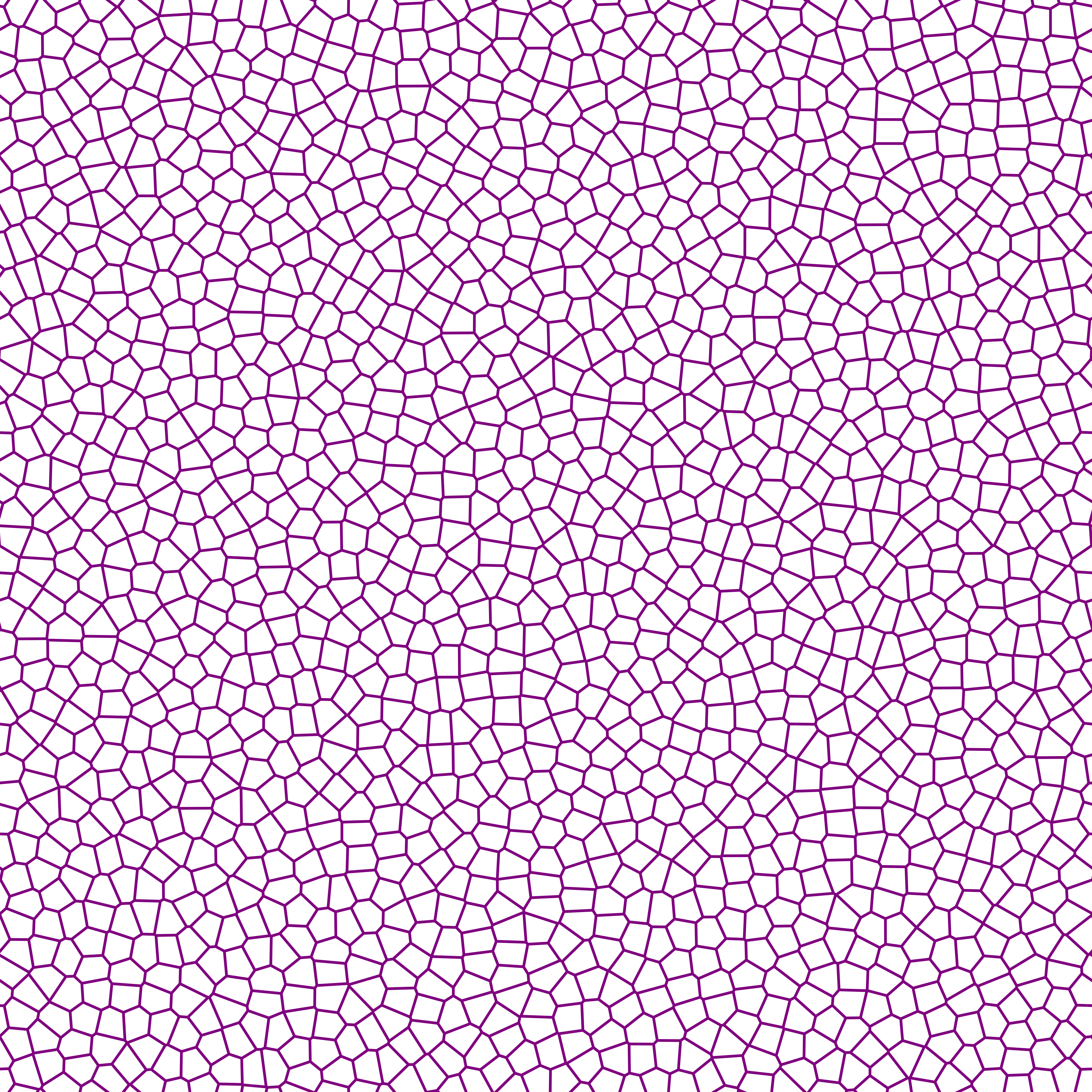}}
        \hspace{0.1\textwidth}
        \subfigure[]{\includegraphics[height=0.4\textwidth]{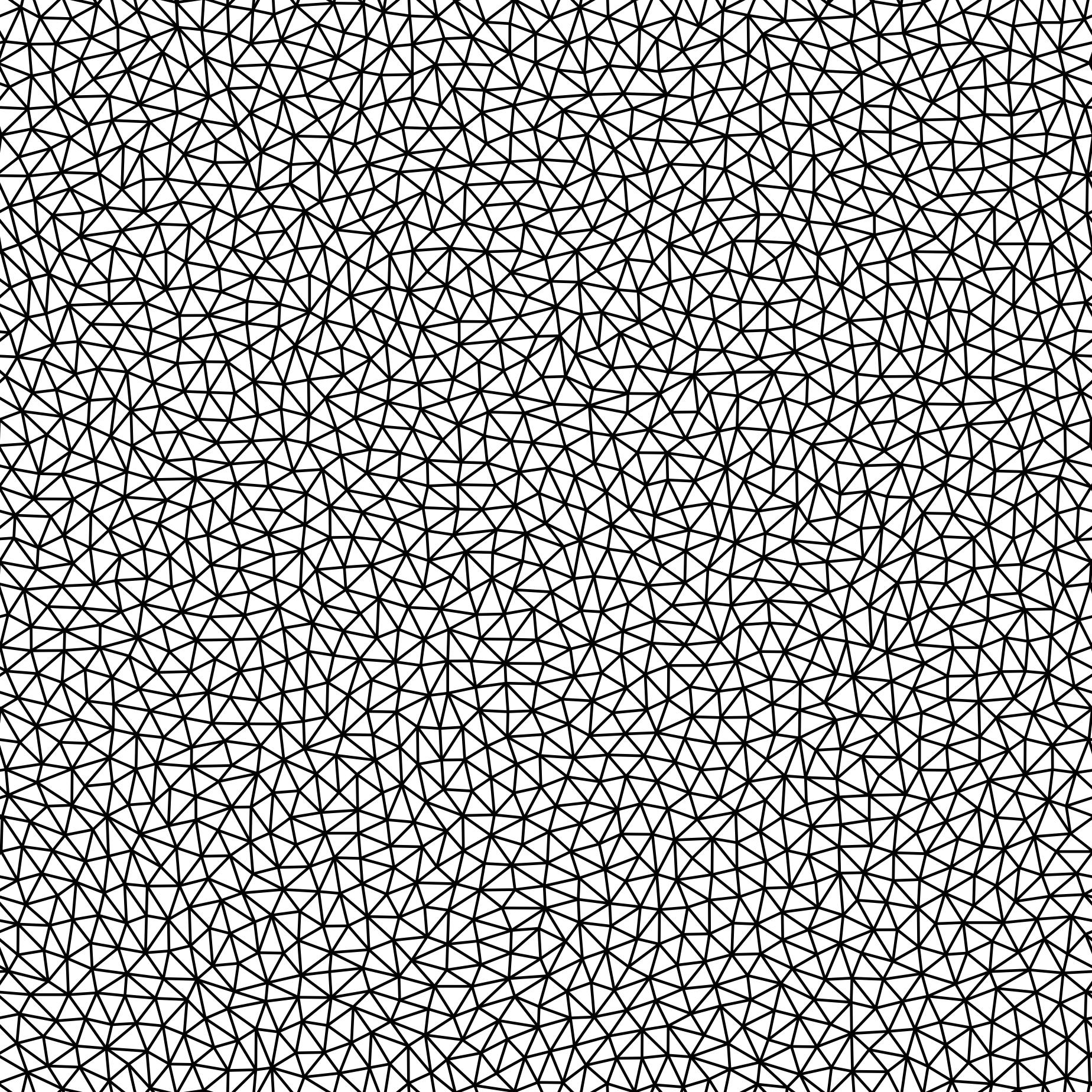}}
        \\
        \subfigure[]{\includegraphics[height=0.4\textwidth]{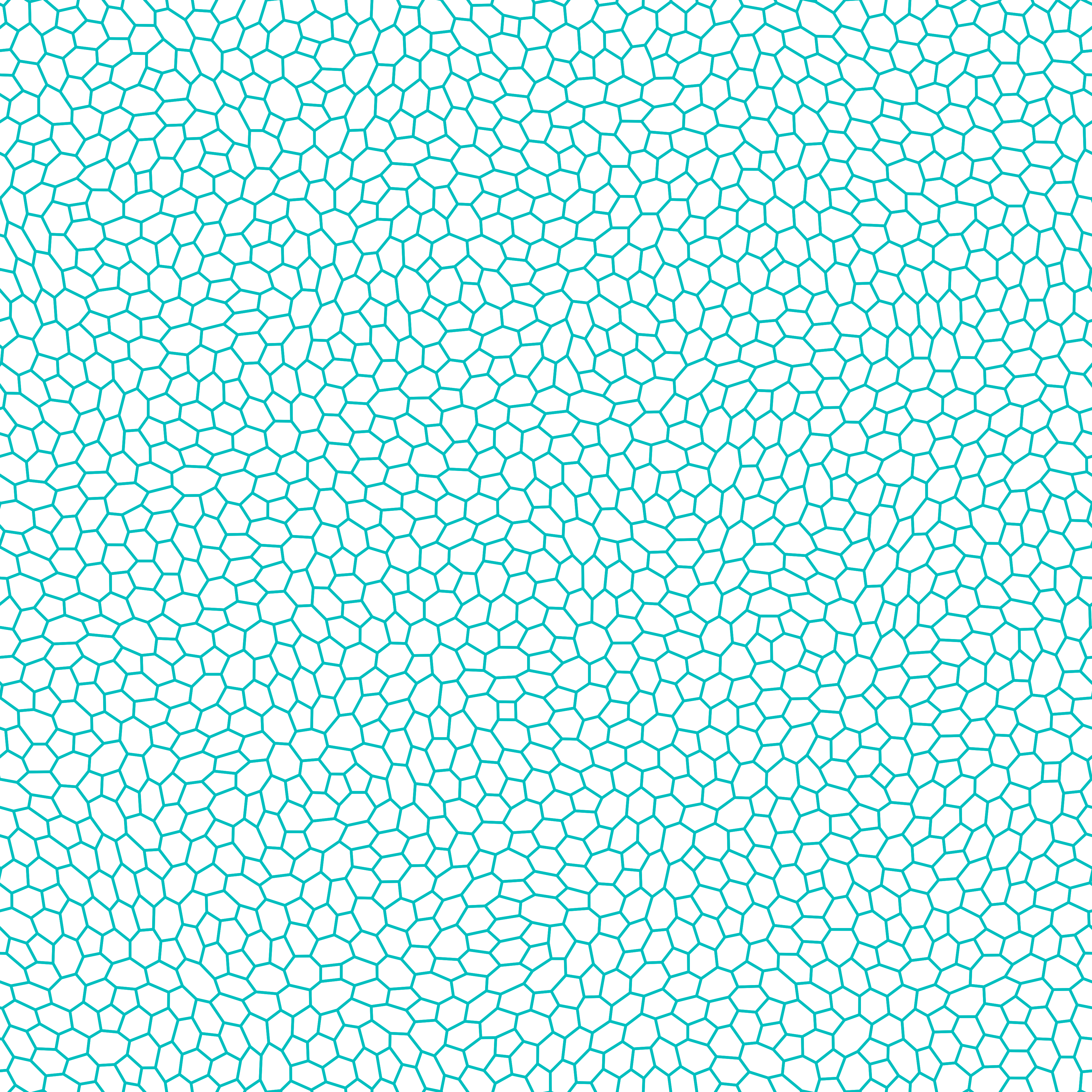}}
        \hspace{0.1\textwidth}
        \subfigure[]{\includegraphics[height=0.4\textwidth]{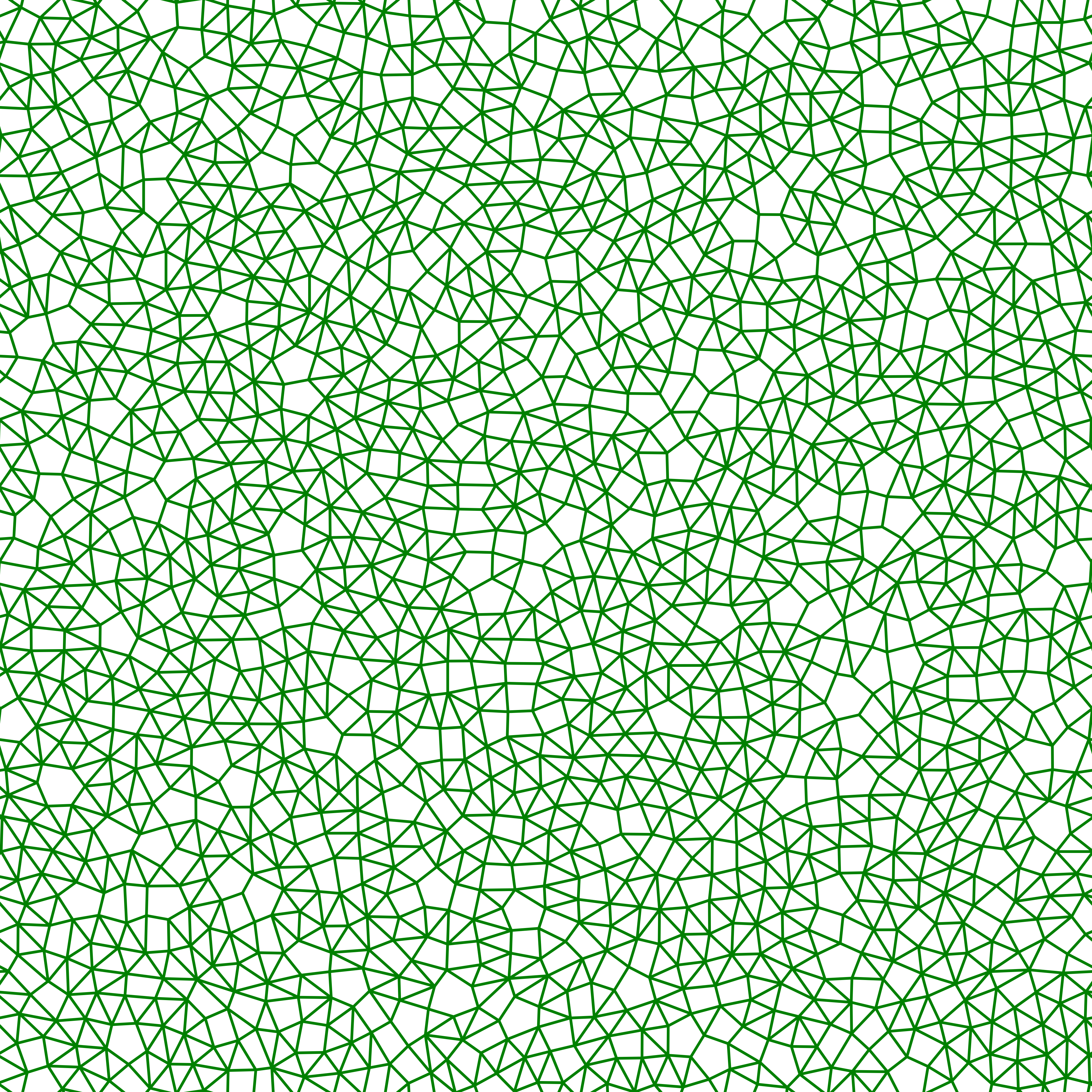}}

        \caption{Example networks derived a (a) Voronoi, (b) Delaunay, (c) Delaunay-centroidal, and (d) Gabriel tessellation of a disordered stealthy hyperuniform point pattern with $\chi = 0.4$.}
    \label{fig:net_sample}
\end{figure*}

\subsection{Hyperuniform and nonhyperuniform model point patterns}
In this work, we consider one nonhyperuniform model and two hyperuniform model point patterns in $d = 2,3$.
The nonhyperuniform model we consider is a totally uncorrelated point process in $\mathbb{R}^d$.
We generate these point patterns at a number density $\rho = 1$ in a fixed $d$-dimensional box by uniformly randomly placing $N$ points independent of each other.
Such a system with no spatial correlations has 
$g_2(r) = S(k) = 1$ and thus is nonhyperuniform (i.e., $\alpha = 0$)  with $\sigma_N^2\sim R^d$ at large $R$.

One hyperuniform model we consider is the uniformly randomized lattice (URL) \cite{klatt_cloaking_2020}.
Here, we make the particular choice of displacing the points of the $\mathbb{Z}^d$ lattice by a random $d$-dimensional vector uniformly distributed on a scaled version of the unit cell of $\mathbb{Z}^d$, i.e., $a\mathcal{F} \equiv [-a/2,a/2)^d$, where $a$ controls the perturbation strength. 
Regardless of the exact value of $a>0$, the URL is a class I hyperuniform system with $\alpha = 2$.
One can tune the degree of local translational disorder in the point pattern with the specific choice of $a$; in particular, larger values of $a$ result in a greater degree of local translational disorder.
Here, we consider values of $a$ between 0.1 and 1.75.

Additionally, we consider disordered stealthy hyperuniform point patterns, which have $S(k) = 0$ for $0 < k \leq K$ and thus are class I hyperuniform.
We generate these point patterns using the collective coordinate optimization scheme starting from totally uncorrelated initial conditions \cite{zhang_ground_2015}. 
The parameter $\chi$ is a dimensionless measure of the ratio of constrained degrees of freedom in the stealthy system (i.e., the number of wavenumbers less than $K$ constrained to be 0) to the total number of degrees of freedom (roughly $dN$, where $N$ is the number of particles in the system).
Relatively unconstrained (low-$\chi$) disordered stealthy systems will have greater translational disorder on short length scales and as $\chi$ increases within the so-called ``disordered regime'' ($0 < \chi < 1/2$ for $d=2,3$ \cite{torquato_ensemble_2015}) the degree of short length scale translational disorder decreases.
In this work, we consider $\chi = 0.2, 0.4, $ and 0.48.

\subsection{Network generation}
To generate the networks examined herein, we first map point patterns under periodic boundary conditions in $\mathbb{R}^2$ ($\mathbb{R}^3$) to sets of polygons (polyhedra) that tile the space using the Voronoi (V), Delaunay (D), Delaunay-centroidal (C), and Gabriel (G) tessellations (described below).
Then, we produce a network from the edges that make up the boundary of each cell in the tessellation.
Each polygonal cell in the Voronoi tessellation is the region of space closer to one point than to any other point in the progenitor point pattern.
The Delaunay tessellation is the dual network of the Voronoi tessellation.
The Delaunay-centroidal tessellation \cite{florescu_designer_2009} is generated by connecting the centroid of each simplex in the Delaunay tesselation to the centroid of each neighboring simplex, i.e., those that share an edge (or face in $\mathbb{R}^3$).
Finally, the Gabriel tessellation \cite{gabriel_new_1969} is derived from the Delaunay tessellation by removing any edge if a sphere generated using that edge as its diameter contains another vertex in the tessellation.

In addition to considering these sets of edges, in $\mathbb{R}^2$ we consider the two-phase medium produced when a rectangle of dimensions $2\delta\times b+2\delta$ is superposed on each edge, where $b$ is the Euclidean length of the original edge and $\delta$ is some prescribed amount by which the edge is ``thickened.''
Here, we choose $\delta$ to be equal to one twentieth of $\rho^{-1/d}$.
We note here that the particular choice of beam thickness only impacts the small-scale structural characteristics of the medium; the large-scale density fluctuations are not significantly impacted.
In Fig. \ref{fig:net_sample} we show an example of each of these types of networks.
Because this mapping of point patterns to two-phase media is a perturbation of the underlying network structure, we additionally desire a method to directly examine the network structure.
We introduce and discuss the merits of such a method in Sec. \ref{Sec:NetworkHU}
\section{Toward a Generalization of Hyperuniformity to Spatial Networks}\label{Sec:NetworkHU}

In the previous section, we described how we can decorate network structures embedded in $\mathbb{R}^2$ to generate two-phase media, so that we can use standard techniques (e.g., the spectral density or volume fraction variance) to characterize their density fluctuations across length scales.
However, treatment of network structures as a set of edges (i.e., without decorations) embedded in $\mathbb{R}^d$ are objects of great interest and have wide-ranging applications from the modeling of public transportation systems to neuron activity in the brain, among many others (see, e.g., Ref. \citenum{barthelemy_spatial_2022} and references therein).
Moreover, for the purposes of the present work, it is not yet known how the decoration scheme chosen here perturbs the density fluctuations of the underlying network structure, so we desire a way to probe the structural characteristics of the network directly.

To carry out this characterization, we propose computing the \textit{edge-length variance} $\sigma^2_{\ell}(R)$, which is the variance of the total edge length of a network embedded in $\mathbb{R}^d$ within a uniformly randomly positioned observation window of radius $R$.
Such a characterization method is motivated by the collection of variance measurements that were recently shown to effectively characterize the translational disorder in point patterns \cite{maher_local_2024}, two-phase media \cite{torquato_local_2022}, and scalar fields \cite{skolnick_quantifying_2024} across length scales.
For each (hyper)spherical sampling window placed into the system, we sum all of the Euclidean lengths of edges that fall entirely within the window. 
For edges that fall partially within the window (i.e., one vertex of the edge is inside the window and the other is outside), we consider only the length of the portion of the edge that lies within the window; such a treatment of edge length is naturally connected to the concept of metric networks \cite{bottcher_dynamical_2024}.
While in this work the edges in our networks are weighted by their Euclidean length, this method can be generalized to treat spatially embedded networks with arbitrary weighting schemes.
In these more general cases, we instead sum the weights of all of the edges that lie entirely within the window, and for edges that fall partially within the window we add a fraction of the weight of that edge proportional to the fraction of the Euclidean edge length that falls within the sampling window.

For the two-phase media generated using the scheme in Sec.~IID, the edge-length variance can be linked to the volume-fraction variance by first noting that the volume fraction $\phi$ of the network edges with a prescribed thickness (see previous section for the 2D construction) in an observation window with radius $R$ is approximately proportional to the total length $\ell$ of the edges in the window. 
Additionally, there is a correction term that accounts for the overlap of the thickened edges. Together, the relationship between volume fraction $\phi$ and total edge length $\ell$ is given by 
\begin{equation}
\phi=\frac{\ell\delta+ O\big(\delta^d N(R)\big)}{v_1(R)} 
\end{equation}
in the limit that the edge thickness $\delta\rightarrow0$, where $N(R)$ is the number of network nodes within the spherical region with radius $R$.  
Thus, the volume fraction variance $\sigma^2_V(R)$ is linked to $\sigma_\ell^2(R)$ via
\begin{equation}
    \sigma_V^2(R)=\frac{\delta^2\sigma^2_\ell(R) + O\big(\delta^{2d} \sigma_N^2(R)\big) }{v_1^2(R)} \quad\textrm{as }\delta\rightarrow0.
\end{equation}
From this, we can see that the large-$R$ scalings of 
$\sigma_V^2(R)$ given in Eq.~\eqref{BiV_classesI} 
correspond to the edge-length variance scalings
\begin{equation}\label{eq:Biell_classes}
    \sigma_\ell^2(R)\sim
    \begin{cases}
        R^{d-1},\quad\alpha > 1 &\textrm{(class I)}\\
        R^{d-1}\textrm{ln}(R),\quad\alpha = 1&\textrm{(class II)}\\
        R^{d-\alpha},\quad0<\alpha<1&\textrm{(class III)},
    \end{cases}
\end{equation}
for hyperuniformity and $R^d$ for nonhyperuniformity.
We note that such scalings correspond to those of the number variance $\sigma_N^2(R)$ for point patterns given in Eq.~\eqref{eq:BiN_classes}.
Deriving the corresponding scattering function for these network structures to supplement $\sigma_\ell^2(R)$ is an important open problem in the field.

While the spectral density methods used in this work are reasonably fast---on the order of several minutes per structure for the system sizes used herein when run on a Windows 11 operating system with 16 GB of RAM and a 12th Gen Intel Core i7-1255U processor running at 1.70 GHz---they suffer from potential discretization errors, especially when generating the pixelizations of very large or nearly crystalline networks.
Moreover, one must choose a particular beam thickness (and shape for $d > 2$), which may impact the density fluctuations of the structure in unexpected ways.
Both of the aforementioned issues can be avoided when computing $\sigma^2_{\ell}(R)$ because the intersections between line segments (the network edges) and hyperspheres (the observation windows of radius $R$) can straightforwardly be computed exactly and no decoration of the network structure is required to do the computations.
The volume-fraction variance can also be computed for some two-phase media without discretization, but computing the overlap of an observation window with such a medium is more computationally intensive than the corresponding overlap calculation for $\sigma^2_{\ell}(R)$. 
Computing $\sigma^2_{\ell}(R)$ for a network structure derived from a $N = 300^2$ 2D point pattern takes on the order of an hour, while the corresponding volume fraction variance calculation for the overlapping rectangle system takes several hours, where both calculations are done using the same computer specifications as above.

\section{Results}\label{sec:Results}

\subsection{Two-dimensional networks}\label{Sec:2DRes}
Here, we present the ensemble-averaged spectral densities $\tilde{\chi}_{_V}(k)$ obtained using Eq. \ref{eq:SpecDen_Pix} on pixelized images of V, D, C, and G networks derived from totally uncorrelated, URL, and disordered stealthy hyperuniform point patterns in $\mathbb{R}^2$.
To compare two-phase structures generated using many combinations of point patterns and tessellation schemes, we use the specific surface $s$---the interfacial area per unit volume---to make length scales dimensionless, as advocated for in Refs. \citenum{torquato_local_2022, kim_characterizing_2021}.

\subsubsection{Totally uncorrelated networks}\label{sec:2DPoi}
\begin{figure}[!t]
    \centering
    \includegraphics[width=0.9\linewidth]{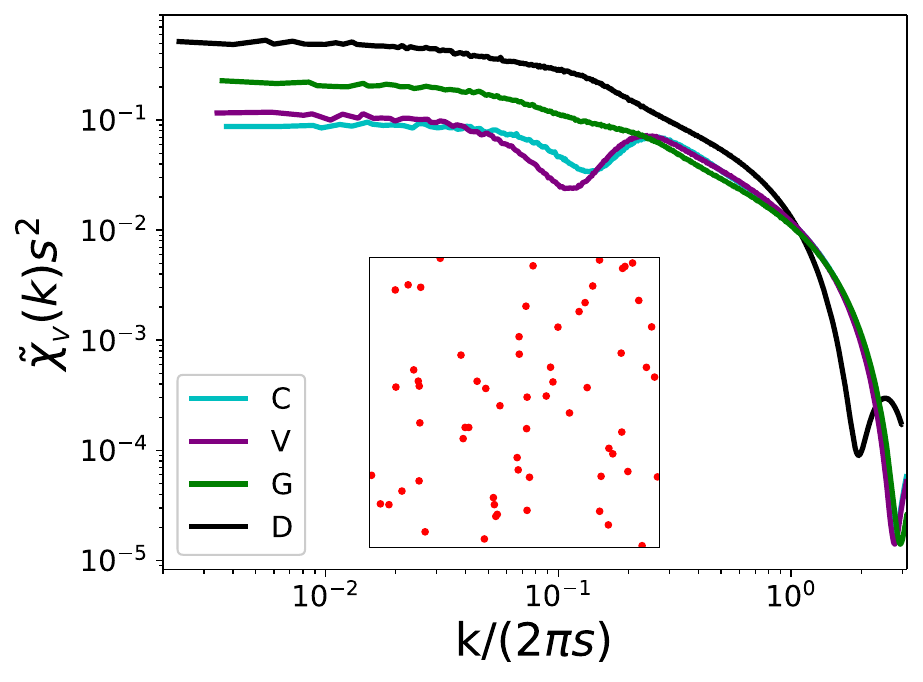}
    \caption{The dimensionless spectral density $\tilde{\chi}_{_V}(k)s^2$ as a function of dimensionless wavenumber $k/(2\pi s)$ for V, D, C, and G networks derived from totally uncorrelated point patterns. The inset shows a small section of a totally uncorrelated point pattern.}
    \label{fig:2D_Poi_Xv}
\end{figure}

Figure \ref{fig:2D_Poi_Xv} shows the dimensionless spectral density $\tilde{\chi}_{_V}(k)s^2$ as a function of dimensionless wavenumber $k/(2\pi s)$ for V, D, C, and G networks derived from totally uncorrelated nonhyperuniform point patterns.
Each curve is an ensemble average of 100 such networks derived from configurations with $N = 100^2$ points.
Noise is present in the small-$k$ region of these curves (and all others in this work) due to statistical fluctuations between different configurations in the same ensemble.

At small $k$, the spectral density of all four types of network flattens out, which indicates that these networks are nonhyperuniform. 
This behavior is consistent with the nonhyperuniformity of the progenitor totally uncorrelated point patterns, indicating that the nonhyperuniformity of point patterns is readily inherited by the corresponding network structures.
While all four network types are clearly nonhyperuniform, the degree to which density fluctuations are suppressed at intermediate and large length scales varies between the different network types, suggesting that the inheritance of density fluctuations from a progenitor point pattern is not the same across tessellation schemes.
In particular, V and C networks have suppressed density fluctuations at intermediate length scales, while D and G networks have density fluctuations that increase and then plateau as the length scale increases.
Moreover, at the largest accessible length scales in our networks we find that D networks have the greatest large-scale density fluctuations, followed by G, V, and C networks in order of decreasing large-scale fluctuations.

\subsubsection{URL networks}\label{sec:URL2D}
\begin{figure}
    \centering
    \subfigure[]{\includegraphics[width=0.9\linewidth]{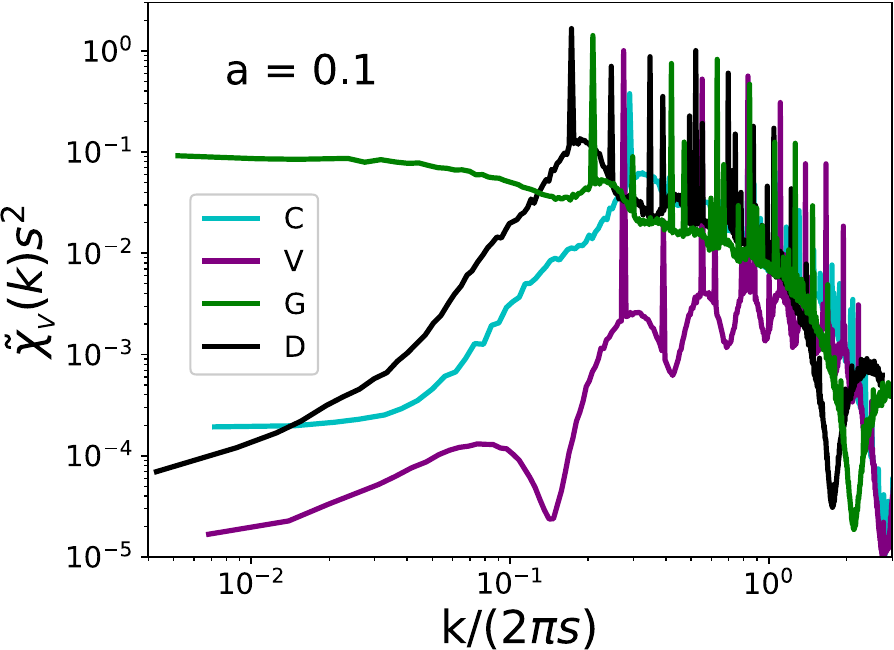}}
    \centering
    \subfigure[]{\includegraphics[width=0.9\linewidth]{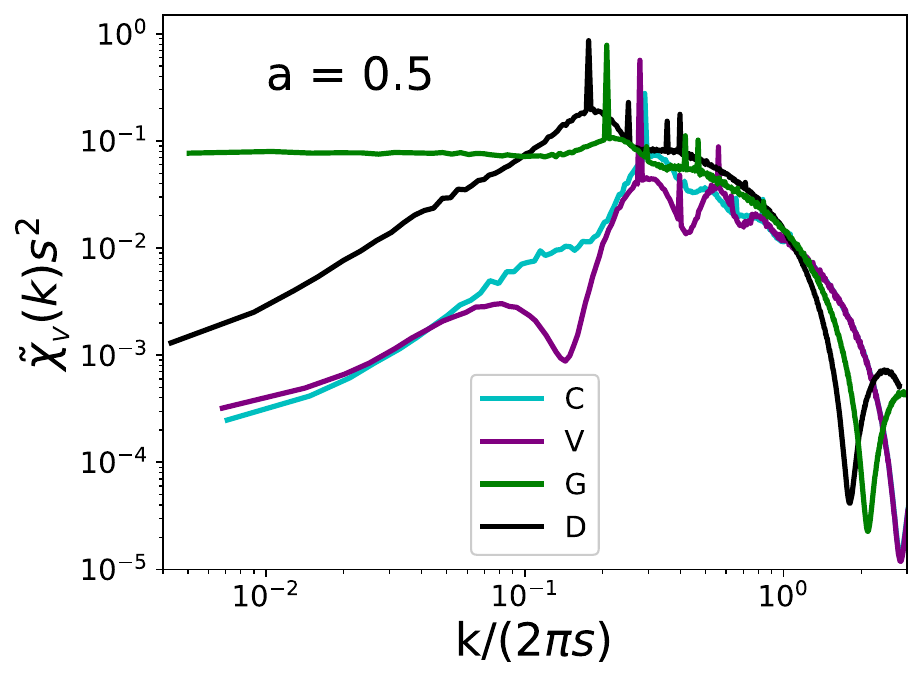}}\\
    \centering
    \subfigure[]{\includegraphics[width=0.90\linewidth]{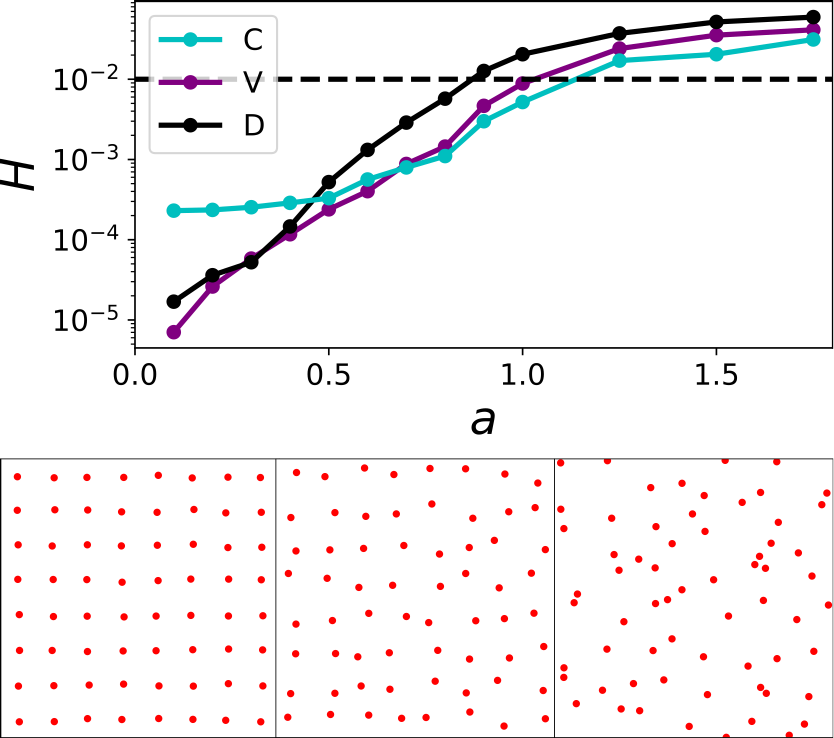}}
    \caption{The dimensionless spectral density $\tilde{\chi}_{_V}(k)s^2$ as a function of dimensionless wavenumber $k/(2\pi s)$ for V, D, C, and G networks derived from URL point patterns with (a) $a = 0.1$ and (b) $a = 0.5$. Panel (c) shows the hyperuniformity index $H$ as a function of the perturbation parameter $a$ for V, D, and C networks derived from URL point patterns with $a\in[0.1,1.75]$. Points that fall below the horizontal black line are effectively hyperuniform ($H\lesssim10^{-2}$), while those above are not. The three lower panels in (c) correspond to small sections of URL point patterns with $a = 0.1$ (left), 0.5 (center), and 1.0 (right).}
    \label{fig:2DURL_Xv}    
\end{figure}

Figure \ref{fig:2DURL_Xv} (a) shows the dimensionless spectral density $\tilde{\chi}_{_V}(k)s^2$ as a function of dimensionless wavenumber $k/(2\pi s)$ for V, D, C, and G networks derived from URL point patterns with $a = 0.1$.
Each curve (and all those discussed in this subsection) is an ensemble average of 100 such networks derived from configurations with $N = 50^2$ points.
For these URL-derived networks with relatively small perturbations, we find that G networks are nonhyperuniform, while the C, D, and V networks are effectively hyperuniform. 
The hyperuniformity index $H$ for each of the C, D, and V network ensembles is
$2.3\times10^{-4}, 1.7\times10^{-5}$, and $7.0\times10^{-6}$, respectively.

\begin{figure}[!t]
    \centering
        \subfigure[]{\includegraphics[height=0.7\linewidth]{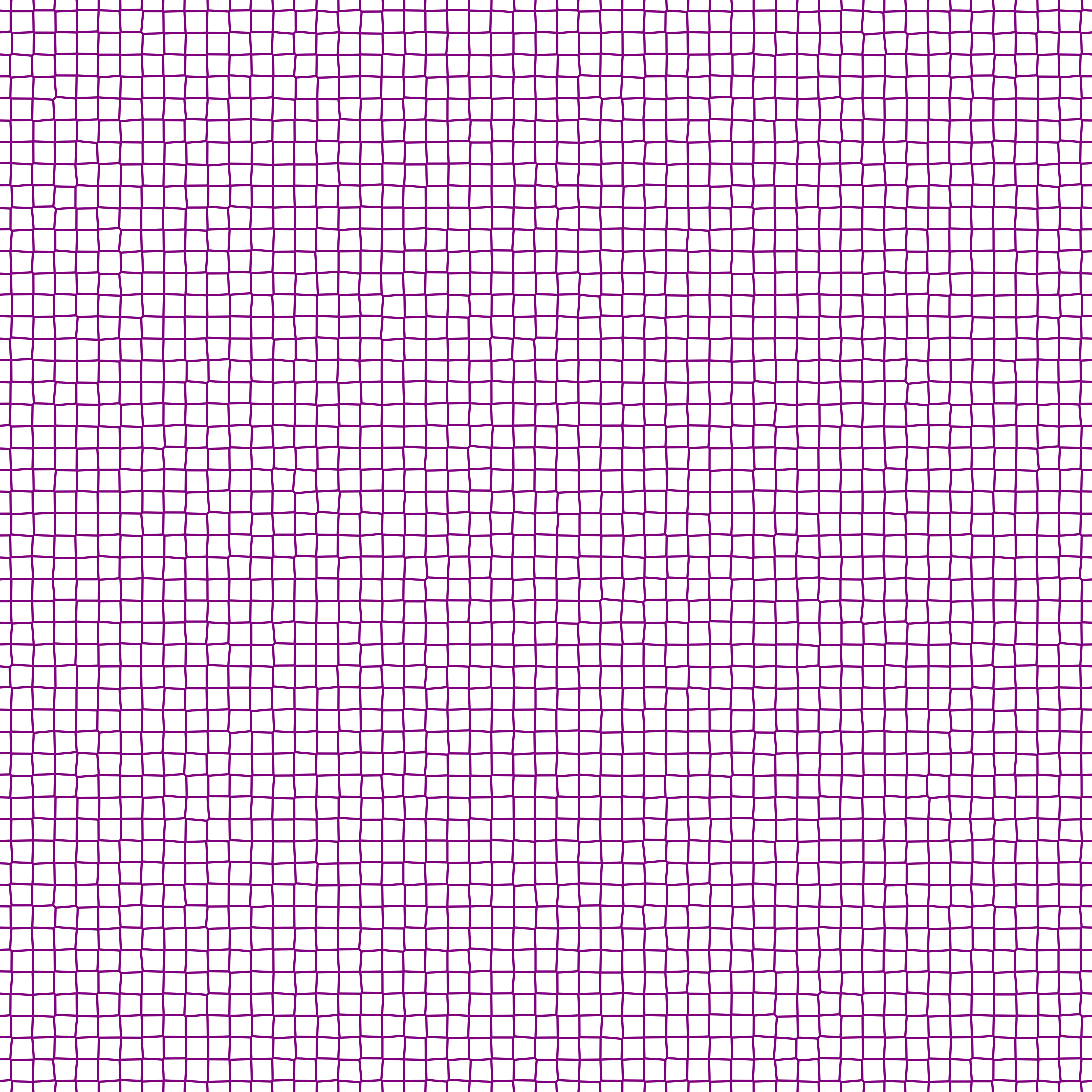}}
        \\
        \subfigure[]{\includegraphics[height=0.7
        \linewidth]{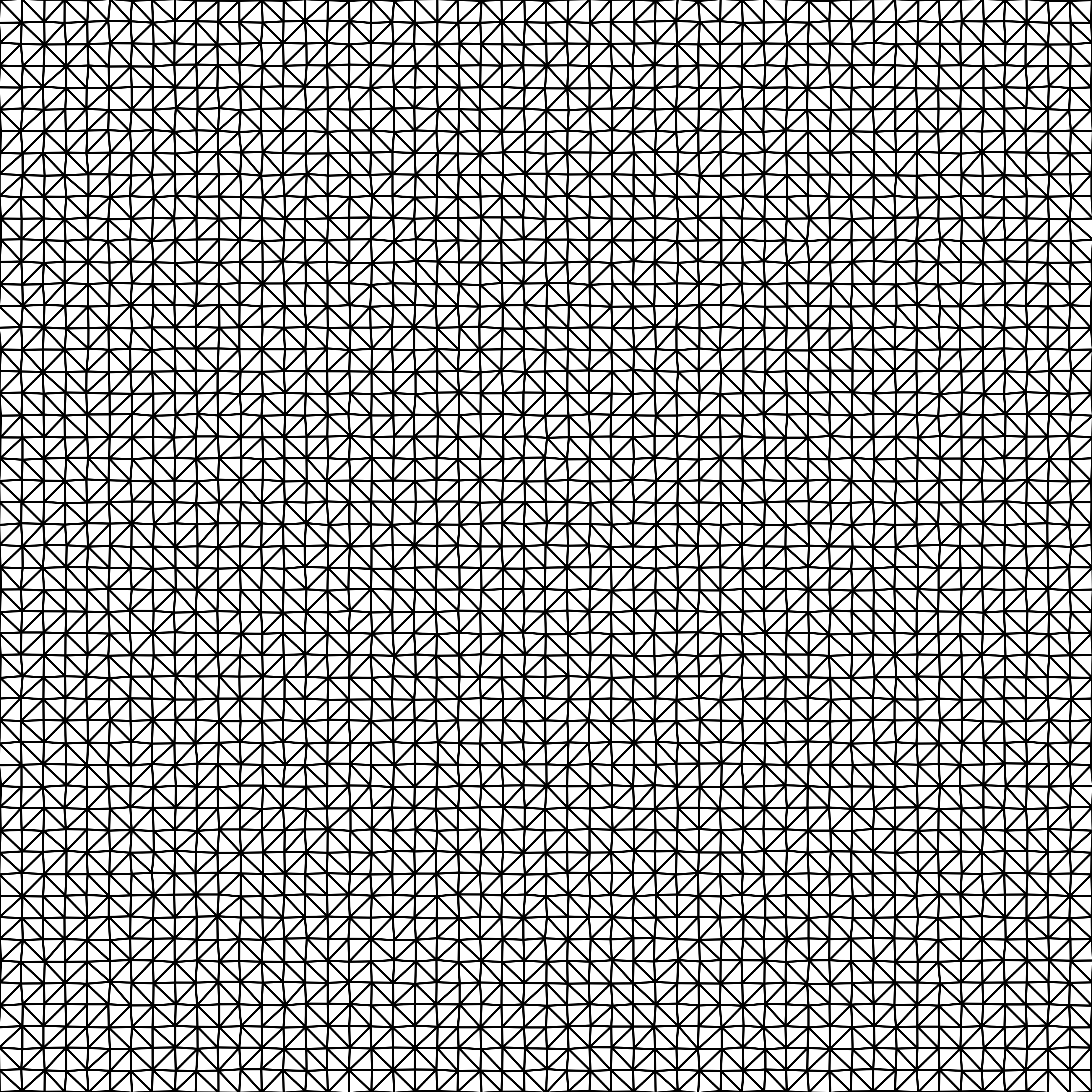}}
        \\
        \subfigure[]{\includegraphics[height=0.7
        \linewidth]{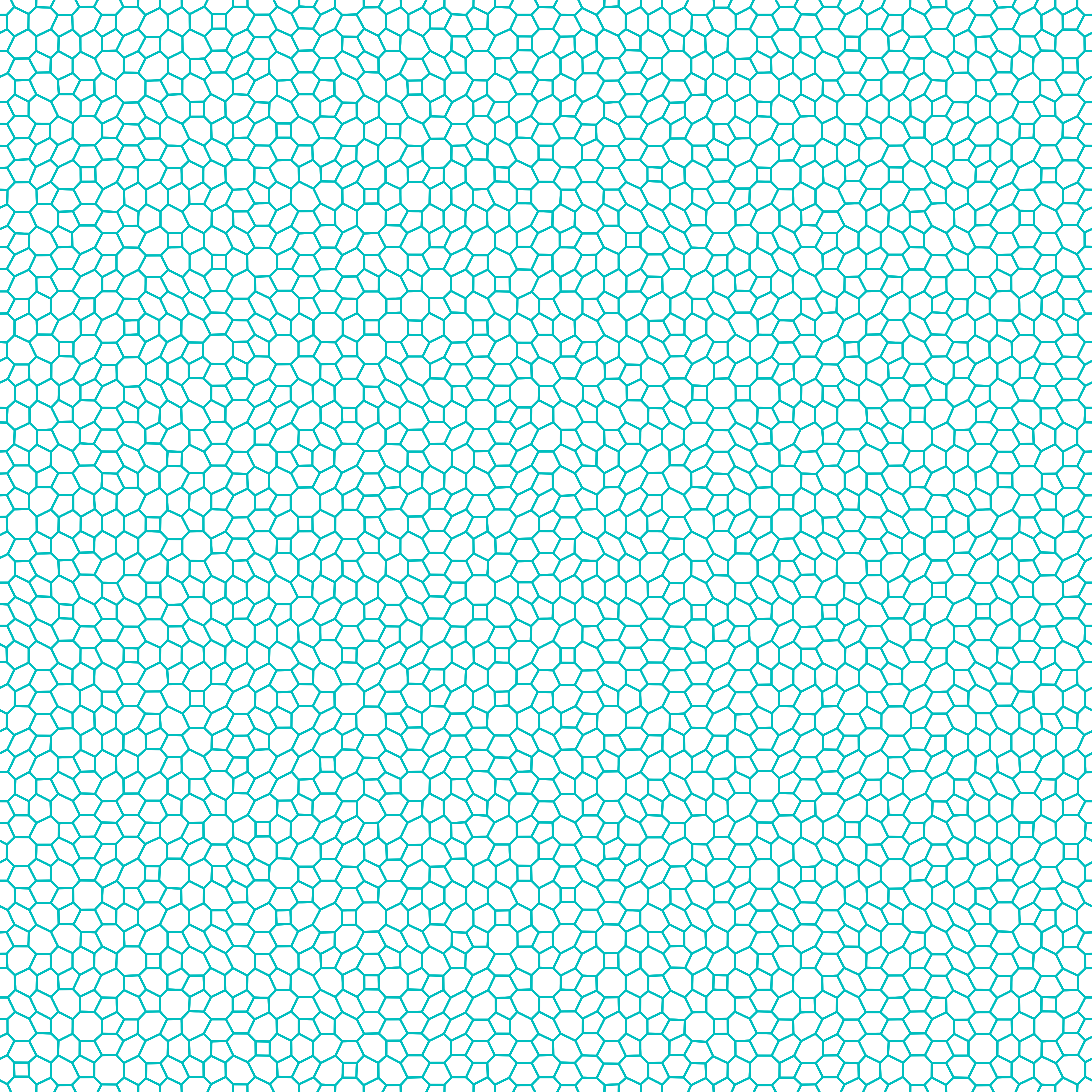}}
        \\

        \caption{Example networks derived from a (a) Voronoi, (b) Delaunay,  and (c) Delaunay-centroidal tessellation of a URL point pattern with $a = 0.1$.}
    \label{fig:smalla_URL_nets}
\end{figure}

This rank-ordering of the effectively hyperuniform small-$a$ URL networks makes intuitive sense because the V networks very closely mimic the V network of the perfect $\mathbb{Z}^2$ lattice.
The D network is slightly more disordered because it is also very close to the V network of the $\mathbb{Z}^2$ lattice but each square cell is split into two triangles along one of its diagonals. 
The C network is the most disordered of these three because it comprises many different possible polygonal cell shapes.
An example of each of these networks is given in Fig. \ref{fig:smalla_URL_nets}.

As $a$ increases above 0.5, we find that the small-$k$ behavior for the C and V networks becomes similar, as observed for the totally uncorrelated point pattern networks, while the D networks tend to fall between the more strongly hyperuniform V and C networks and the nonhyperuniform G networks.
These behaviors are evident in Fig. \ref{fig:2DURL_Xv} (b), which shows the dimensionless spectral density $\tilde{\chi}_{_V}(k)s^2$ as a function of dimensionless wavenumber $k/(2\pi s)$ for V, D, C, and G networks derived from URL point patterns with $a = 0.5$.
In Fig. \ref{fig:2DURL_Xv} (c) we show the values of $H$ as a function of $a$ for the V, D, and C networks.
Here, we see that at small $a$ the hyperuniformity index for the D and V networks increases much more quickly than for the C networks, and at $a = 0.5$ the rank-ordering of degree of hyperuniformity switches from V $>$ D $>$ C to V $>$ C $>$ D.
This rank-ordering switches again at $a = 0.7$ to C $>$ V $>$ D and persists up to the largest values of $a$ examined in this work. 
We expect this rank-ordering to persist as $a \rightarrow \infty$.
We also find that the D networks lose effective hyperuniformity between $a = 0.8$ and 0.9 while both the C and V networks lose effective hyperuniformity between $a = 1.0$ and 1.25.

\begin{figure*}[!t]
   
    \subfigure[]{\includegraphics[height=0.23\textheight]{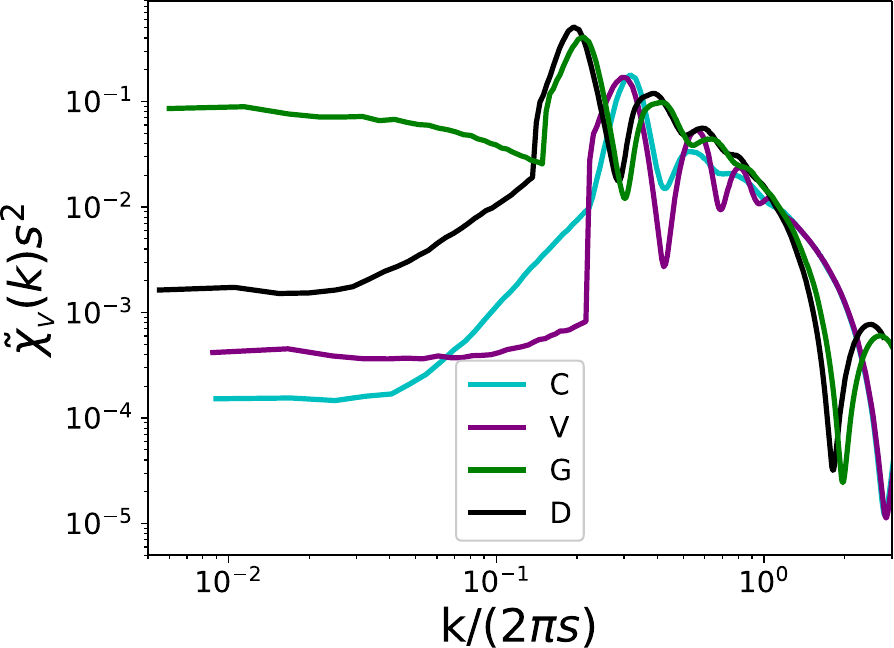}}
    \subfigure[]{\includegraphics[height=0.23\textheight]{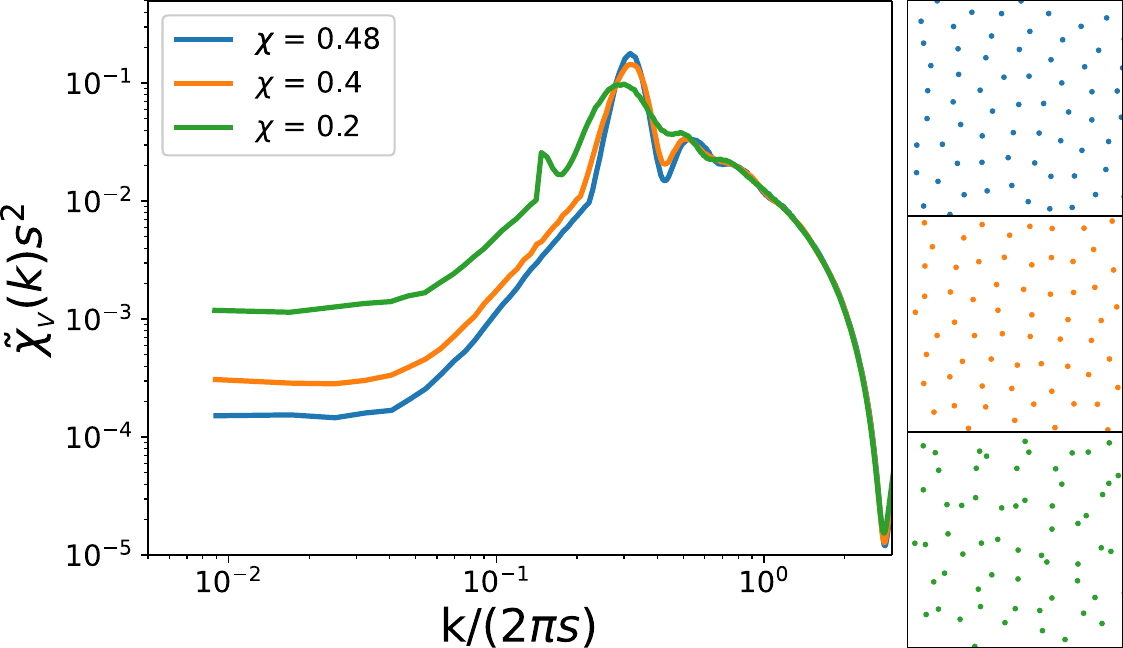}}
    \caption{The dimensionless spectral density $\tilde{\chi}_{_V}(k)s^2$ as a function of dimensionless wavenumber $k/(2\pi s)$ for (a) V, D, C, and G networks derived from disordered stealthy hyperuniform point patterns with $\chi = 0.48$ and (b) for C networks derived from disordered stealthy hyperuniform point patterns with $\chi = 0.48, 0.4$ and 0.2. Panels on the right in  (b) correspond to small sections of a disordered stealthy hyperuniform point pattern with (top) $\chi = 0.48$, (middle) $\chi = 0.4$, and (bottom) $\chi = 0.2$.}
    \label{fig:2DSteal_Xv}
\end{figure*}

\subsubsection{Disordered stealthy networks}
In Fig. \ref{fig:2DSteal_Xv} (a) we show the dimensionless spectral densities $\tilde{\chi}_{_V}(k)s^2$ as a function of dimensionless wavenumber $k/(2\pi s)$ for V, D, C, and G networks derived from ``high-$\chi$'' disordered stealthy hyperuniform point patterns with $\chi = 0.48$.
Each curve in this subsection is an ensemble average of 98 networks derived from configurations with $N = 40^2$ points.
The rank-ordering of hyperuniformity degree in Fig. \ref{fig:2DSteal_Xv} (a) matches the $a > 0.7$ URL network results, i.e., G networks are nonhyperuniform, while C, V, and D networks are effectively hyperuniform (in order of increasing $H$).
In numerically generated disordered stealthy point patterns, the value of $S(k)$ in the ``stealthy region'' ($0<k\leq K$) is not identically zero, but some very small number ($\approx10^{-22}$ in our configurations).
Notably, we find that the shape of the ``stealthy region'' one expects at small-$k$ is present in each of the effectively hyperuniform spectral densities, but at much larger ordinate values compared to the progenitor point pattern.

Additionally, we find that as $\chi$ decreases (i.e., the small length scale translational disorder of the point patterns increases) $H$ increases, analogous to the behavior of $H$ as $a$ increased for the URL-derived networks.
For example, see Fig. \ref{fig:2DSteal_Xv} (b), which shows that the peak heights decrease and the magnitude of the small-$k$ behavior increases as $\chi$ decreases leading to an increase in $H$.
In particular, $H = 8.4\times10^{-4}, 2.1\times10^{-3}, 1.2\times10^{-2}$ for $\chi = 0.48, 0.4, 0.2$, respectively, demonstrating that these C networks lose effective hyperuniformity for $\chi\lesssim0.2$.
The V and D networks lose effective hyperuniformity at larger values of $\chi$ due to their relatively larger values of $H$ than the C networks.

From these results, and those in Secs. \ref{sec:2DPoi} and \ref{sec:URL2D}, it is clear that the degree to which a network structure inherits the hyperuniformity of its progenitor point pattern depends both on the tessellation scheme and the translational order across length scales of the progenitor point pattern.
In particular, while D networks derived from totally uncorrelated point patterns have the greatest density fluctuations at intermediate and large length scales, this is not the case for the URL and stealthy point patterns because D networks are able to partially inherit the hyperuniformity of those point patterns, while G networks do not.
Thus, across networks derived from hyperuniform point patterns, G networks have the greatest large-scale density fluctuations.

Despite the ability of D, V, and C tessellation schemes to inherit some degree of hyperuniformity from their progenitor point pattern, in no case is the underlying hyperuniformity completely inherited.
Moreover, the degree of preservation appears to be impacted by the degree of \textit{local} disorder in the progenitor point pattern, despite hyperuniformity concerning large length scales.
This is best shown by Fig. \ref{fig:2DSteal_Xv} (b), where each of the underlying point patterns is stealthy hyperuniform, but the resulting small-$k$ behavior is different for different $\chi$ values, despite being generated by the same tessellation scheme. 

Another important trend is evident in Fig. \ref{fig:2DURL_Xv} (c), which demonstrates that some tessellation schemes are better at preserving the hyperuniformity of disordered hyperuniform systems with greater degrees of local translational order than others.
In particular, V tessellations better preserve the hyperuniformity of disordered hyperuniform systems with greater degrees of local translational order, while C tessellations better preserve the hyperuniformity of those with lesser degrees of local translational order.
The sensitivity of the rank-ordering of the degree of hyperuniformity to the tessellation and point pattern type is consistent with changes observed in the rank-ordering of certain physical properties across networks generated via different tessellations of different point patterns \cite{torquato_multifunctional_2018}.
 
\subsection{Edge-length variance $\sigma^2_{\ell}(R)$}\label{sec:sigell}
Here, we apply the edge-length variance $\sigma_\ell^2(R)$ to our spatial network structures, to probe their density fluctuations without converting them into two-phase media.
In analogy with the use of the specific surface $s$ to scale distances in two-phase media, we use $\sqrt[d-1]{\rho_{\ell}}$, where $\rho_{\ell}$ is the edge length per unit volume, or the edge-length density, to make distances dimensionless in the analysis of our networks.
Moreover, we note that we use a maximum window radius of $L/4$, where $L$ is the side length of the square (cubic) simulation box, to mitigate finite-size effects.


\begin{figure}[!t]
    \centering
        \subfigure[]{\includegraphics[height=0.3\textwidth]{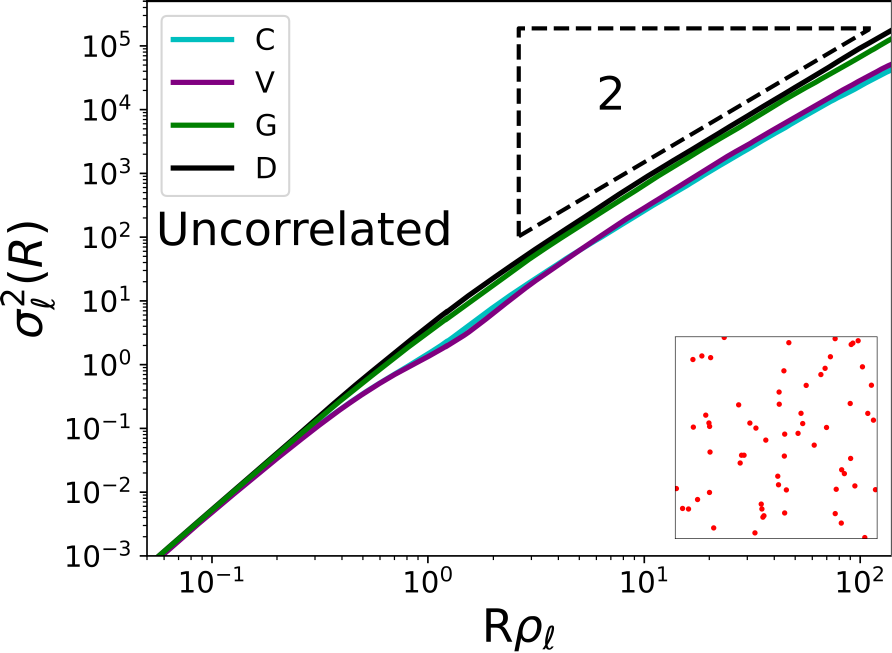}}
        \subfigure[]{\includegraphics[height=0.3\textwidth]{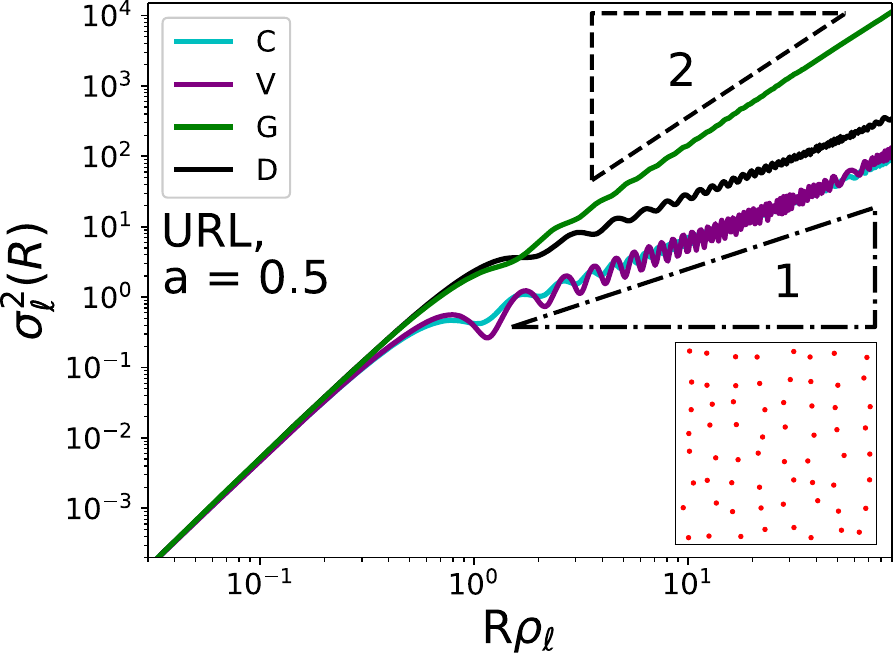}}
        \subfigure[]{\includegraphics[height=0.3\textwidth]{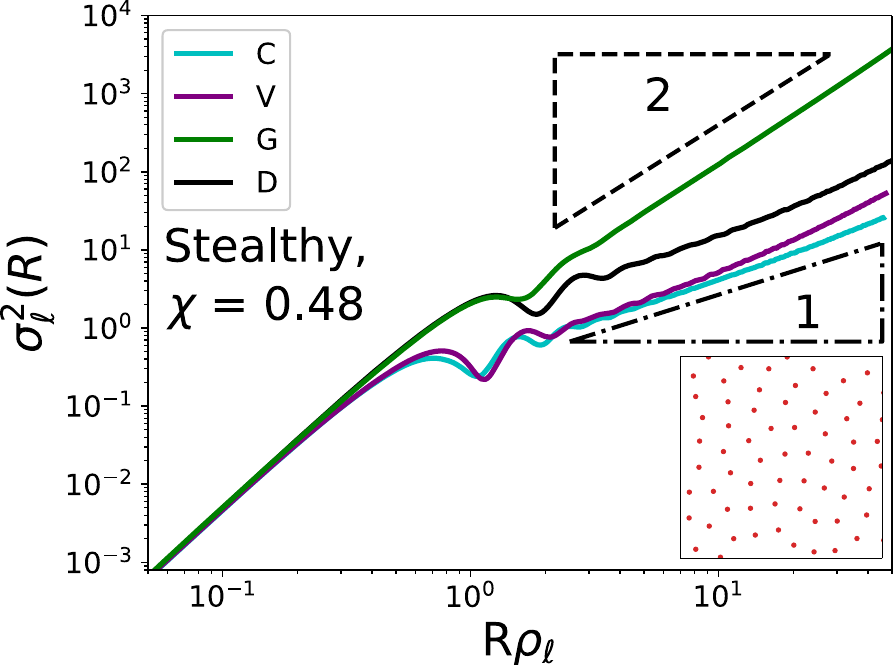}}
    \caption{The edge-length variance $\sigma_{\ell}^2(R)$ as a function of the dimensionless observation window radius $R\rho_{\ell}$ for V, D, C, and G networks derived from (a) totally uncorrelated point patterns (b) URL point patterns with $a = 0.5$ and (c) disordered stealthy hyperuniform point patterns with $\chi = 0.48$. The dashed and dash-dotted triangles are power law eye-guides corresponding to $R^2$ and $R$ scalings, respectively. The insets correspond to small sections of (a) totally uncorrelated, (b) URL with $a$ = 0.5, and (c) stealthy hyperuniform with $\chi = 0.48$ point patterns. }
    \label{fig:2D_ELV}
\end{figure}

In Figs. \ref{fig:2D_ELV} (a) and (b), we show the $\sigma^2_{\ell}(R)$ curves for 50-configuration ensembles of V, D, C, and G networks derived from totally uncorrelated point patterns with $N = 300^2$ and 50-configuration ensembles of V, D, C, and G networks derived from URL configurations with $a = 0.5$ and $N = 200^2$.
In Fig. \ref{fig:2D_ELV} (c) the curves correspond to 98-configuration ensembles of V, D, C, and G networks derived from disordered stealthy hyperuniform point patterns with $\chi = 0.48$ and $N=100^2$.
In each subfigure of Fig. \ref{fig:2D_ELV}, the small-$R$ behavior of each set of curves collapses, indicating that $\sqrt[d-1]{\rho_{\ell}}$ is indeed a reasonable choice of scale.
From each of the four curves in Fig. \ref{fig:2D_ELV} (a) and the G curves in Figs. \ref{fig:2D_ELV} (b) and (c), it is evident that the nonhyperuniform systems examined in Sec. \ref{Sec:2DRes} have $\sigma^2_{\ell}(R)\sim R^2$ (i.e., scale like the observation window volume) at large $R$.
Like in Fig. \ref{fig:2D_Poi_Xv}, one can clearly observe that the ordering of the totally uncorrelated point pattern networks from largest to smallest large-scale density fluctuations is D $>$ G $>$ V $>$ C.
Moreover, the V, D, and C curves in Figs. \ref{fig:2D_ELV} (b) and (c) exhibit $\sigma^2_{\ell}(R)\sim R$ scaling at intermediate length scales, which increases at larger length scales.
This change in scaling behavior is consistent with the onset of degraded hyperuniform scaling at small $k$ evident in Figs. \ref{fig:2DURL_Xv} (b) and \ref{fig:2DSteal_Xv} (a).

From these results, we have confirmed the scaling behavior of $\sigma^2_{\ell}(R)$ given in Eq.~\eqref{eq:Biell_classes} that matches the number variance $\sigma_N^2(R)$ scalings, i.e., nonhyperuniform systems have $R^d$ scaling while hyperuniform systems have slower scaling.
These results also indicate that the onset of the degradation of hyperuniformity observed at large length scales in Sec. \ref{Sec:2DRes} is not due to mapping these networks to two-phase media, but intrinsic to the network structures themselves.
Importantly, this implies that the particular choice of parameters used to convert the network to a two-phase medium does not significantly impact the hyperuniformity of the resulting structure and one can probe the network structure directly to assess the hyperuniformity of structures derived from it.

\subsection{Three-dimensional networks}
Having established that $\sigma^2_{\ell}(R)$ faithfully captures the intermediate and large-scale density fluctuations in 2D network structures, we now apply this same method to 3D networks.
We opt not to map the 3D networks to two-phase media and characterize the resulting structures with either the volume-fraction variance or spectral density due to the significantly increased difficulty of exactly describing the geometry of the media at the network vertices, where several of these beams overlap compared to $d = 2$.
Calculations involving the newly proposed edge-length variance are much easier to execute, by comparison, for the reasons stated in Sec. \ref{Sec:NetworkHU}.

\begin{figure}[b!]
    \centering
    \subfigure[]{\includegraphics[width=0.4\textwidth]{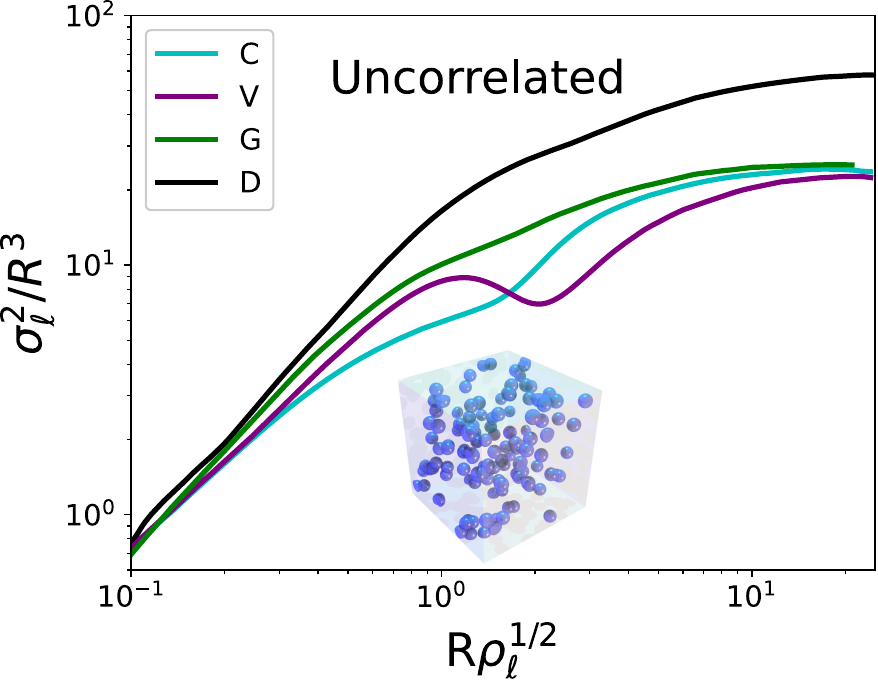}}\\
    \subfigure[]{\includegraphics[width=0.4\textwidth]{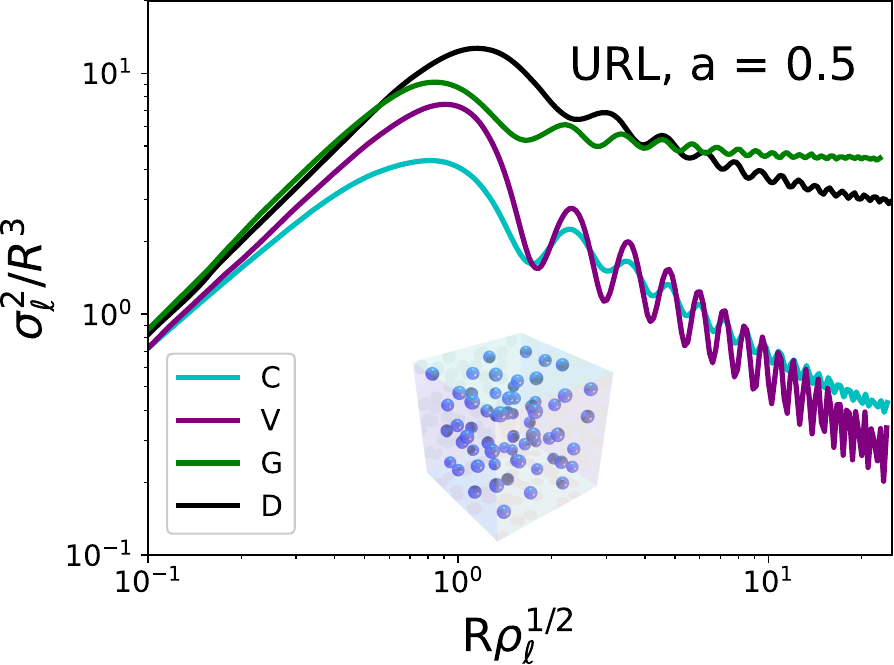}}\\
    \subfigure[]{\includegraphics[width=0.4\textwidth]{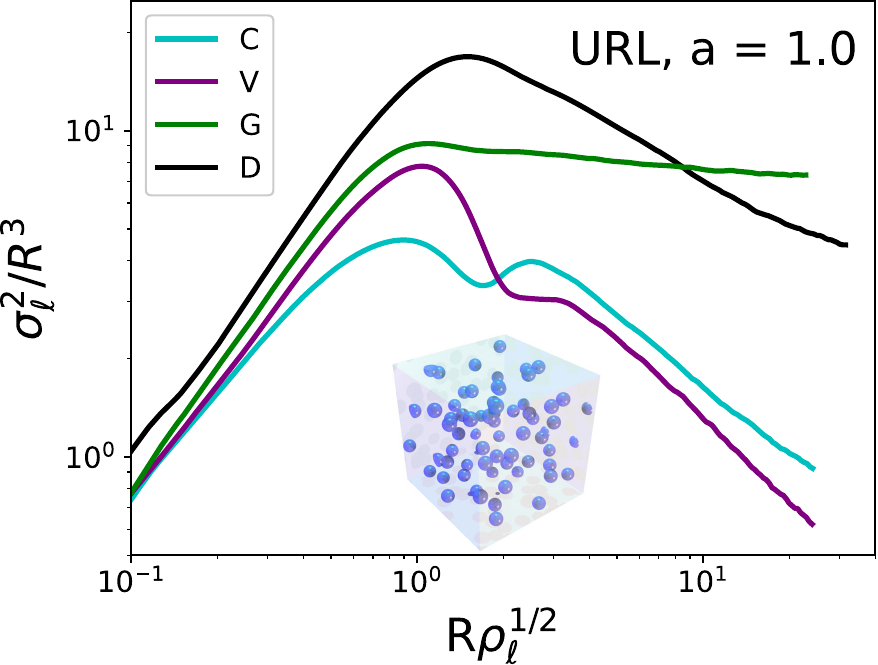}}
    \caption{The edge-length variance scaled by the window volume $\sigma_{\ell}^2(R)/R^3$ as a function of the dimensionless window radius $R\rho_{\ell}^{1/2}$ for V, D, C, and G networks derived from (a) totally uncorrelated point patterns, (b) URL point patterns with $a = 0.5$, and (c) URL point patterns with $a = 1.0$. Insets correspond to small sections of (a) totally uncorrelated, (b) URL with $a = 0.5$, and (c) URL with $a = 1.0$ point patterns.
    }

    \label{fig:ELV_3D}
\end{figure}

In Fig. \ref{fig:ELV_3D} (a) we present the edge-length variance scaled by the window volume $\sigma_{\ell}^2(R)/R^3$ as a function of the scaled window radius $R\rho_{\ell}^{1/2}$ for 500-configuration ensembles of V, D, C, and G networks derived from 3D totally uncorrelated point patterns with $N = 40^3$ points.
Here, we scale $\sigma_{\ell}^2(R)$ by the observation window volume $R^3$ for the sake of visual clarity. 
Curves that plateau at large $R$ correspond to nonhyperuniform structures and those that decrease at large $R$ correspond to hyperuniform structures.
Figure \ref{fig:ELV_3D} (a) reveals that the networks derived from totally uncorrelated point patterns have nonhyperuniform $\sigma_{\ell}^2(R)\sim R^3$ scaling consistent with the nonhyperuniformity of the progenitor point pattern.
However, the rank-ordering of large-scale density fluctuations in these totally uncorrelated networks changes from D $>$ G $>$ V $>$ C for the $d=2$ case (cf. Fig. \ref{fig:2D_ELV} (a)), to D $>$ G $>$ C $>$ V.

Figures \ref{fig:ELV_3D} (b), and (c) show the edge-length variance scaled by the window volume $\sigma_{\ell}^2(R)/R^3$ as a function of the scaled window radius $\sigma^2_{\ell}(R)$ for 500-configuration ensembles of V, D, C, and G networks derived from 3D URL point patterns with $N = 40^3$ points and $a = 0.5$ and $1.0$, respectively.
In these figures, we observe the expected behavior of the G networks, i.e., they have more obvious nonhyperuniform scaling than the V, C, and D networks and the greatest density fluctuations at large $R$.
The rank-ordering of the other three network types for stealthy and large-$a$ URL point patterns, however, is different from the 2D case.
Specifically, while D is still the most weakly hyperuniform of the three, V networks are consistently more hyperuniform than the corresponding C networks at large $R$.
The change observed here in the rank-ordering between $d = 2$ and $d = 3$ is not surprising considering that the structures that minimize large-scale density fluctuations in point patterns also change across dimensions (e.g., the $A_2$ lattice minimizes such fluctuations in $d=2$, while $A_3^*$ is conjectured to do so in $d = 3$ \cite{torquato_local_2003}).
We also note that, on intermediate length scales, D networks appear to have the greatest density fluctuations of the four network types, regardless of the (non)hyperuniformity of the progenitor point pattern.
In addition, while the onset of the degraded hyperuniform scaling at large length scales is visible in \ref{fig:ELV_3D} (b), it is less obvious in \ref{fig:ELV_3D} (c), suggesting that the degradation of hyperuniformity occurs at larger length scales in systems with greater degrees of disorder on small length scales.

\section{Discussion and Conclusions}\label{sec:Conc}
In this work, we probed the ability of the first step of the configuration-to-network-to-3D printed structure pipeline devised by Obrero et al. \cite{obrero_electrical_2024} to generate hyperuniform network structures.
In particular, we generated Voronoi, Delaunay, Delaunay-Centroidal, and Gabriel tessellations of totally uncorrelated, uniformly randomized lattice, and disordered stealthy hyperuniform point patterns in $\mathbb{R}^2$ and $\mathbb{R}^3$.
We then characterized the structures of the networks generated by the connected sets of edges in the aforementioned tessellations in two ways.
First, for networks in $\mathbb{R}^2$, we created two-phase media by mapping each edge in the network to a rectangle and then computed the spectral density $\tilde{\chi}_{_V}(k)$ of the resulting structure.
We then proposed a new variance-based measure inspired by other recent variance-based characterizations of disorder across length scales \cite{torquato_local_2022, skolnick_quantifying_2024, maher_local_2024}, the edge-length variance $\sigma_{\ell}^2(R)$, to characterize the structures of spatially embedded networks directly without needing to convert to structure to a two-phase medium.
We also suggested the use of the $(d-1)$th root of the edge-length density $\rho_{\ell}$ to make length scales dimensionless when comparing different types of networks.
We then applied this new characterization method to the networks derived from the aforementioned point patterns in both $\mathbb{R}^2$ and $\mathbb{R}^3$.

By examining the spectral densities of two-phase media derived from 2D networks, we found that networks derived from nonhyperuniform point patterns inherit the nonhyperuniformity of the progenitor point pattern, but tessellations of hyperuniform point patterns only partially inherit the hyperuniformity of the progenitor point pattern (or not at all, in the case of G tessellations).
We found that the C, V, and D  
network structures are, at best, effectively hyperuniform and in other cases nonhyperuniform.
For disordered hyperuniform systems with a large degree of local translational order, V networks best preserve hyperuniformity, while C networks best preserve the hyperuniformity of disordered hyperuniform point patterns with a small degree of local translational order.
We also found that, despite hyperuniformity describing the large-scale structure of a system, a network derived from a hyperuniform point pattern with a small degree of local translational order will be more weakly hyperuniform than one derived from a point pattern that has the same strength of hyperuniformity but has a greater degree of local translational order.
Moreover, the rank-ordering of the degree of hyperuniformity in network structures is sensitive to the particular tessellation scheme and both the small- and large-scale translational order of the progenitor point pattern.
Such a sensitivity is consistent with the sensitivity of the rank-ordering of the transport and elastic properties of these network structures to the same variables \cite{torquato_multifunctional_2018}. 

We then showed that the edge-length variance $\sigma_{\ell}^2(R)$ faithfully captures the intermediate- and large-scale structural characteristics of the networks when treated as a set of line segments (i.e., not treated as a two-phase medium) and has the same $R^d$ scaling for nonhyperuniform systems and slower scaling for hyperuniform systems as the number variance $\sigma_N^2(R)$.
The calculation of $\sigma_{\ell}^2(R)$ is much faster and simpler compared to the corresponding spectral density or volume-fraction variance calculations due to the relative simplicity of the line-segment-based network structure.
Moreover, we showed that the degradation of hyperuniformity at large length scales observed in the spectral densities of V, D, and C networks derived from 2D hyperuniform point patterns is also present in the $\sigma_{\ell}^2(R)$ curves, suggesting that this degradation of hyperuniformity is intrinsic to the networks structures.
This finding suggests that the particular choice of beam shape when designing two-phase structures for manufacturing purposes will not significantly impact the large-scale density fluctuations of the manufactured structure---an important practical consideration.
We then applied $\sigma_{\ell}^2(R)$ to 3D networks derived from nonhyperuniform and disordered hyperuniform point patterns and found that V networks tend to be the best at suppressing large-scale density fluctuations.
This change in optimal structure from C or V networks for $d = 2$ to $V$ for $d = 3$ is not unexpected given that the currently known structures that minimize large-scale density fluctuations in point patterns also change across space dimensions \cite{torquato_local_2003}.

We note here that, while we have only run numerical experiments to examine the phenomena discussed in this paper, we are reasonably confident that the behaviors we observe in the large-scale density fluctuations of our network structures are not numerical artifacts.
However, this lack of theoretical background motivates the study of the link between some spectral function and the edge-length variance (like the link between the spectral density and the volume fraction variance).
Such a link would be of great interest to both the metamaterial and network science communities.

The results presented here can be used to inform the design of effectively hyperuniform network metamaterials in two and three space dimensions for future theoretical and experimental characterization of their physical properties, e.g., their thermal conductivity, electrical conductivity, and acoustic properties.
Another potentially interesting area of future research is the application of $\sigma_{\ell}^2(R)$ to existing network structures that have been shown to possess desirable physical properties, e.g., the disordered 2D networks with hyperuniform vertex locations described in Refs. \citenum{chen_stonewales_2021} and \citenum{chen_topological_2021}, or the naturally occurring looped leaf vein structures discussed in Ref. \cite{liu_universal_2024}.
Moreover, our results demonstrating the effective hyperuniformity or nonhyperuniformity of networks derived from the spatial tessellations of various hyperuniform point patterns on large length scales motivate a search for network structures that are truly hyperuniform as opposed to effectively hyperuniform.
Possible inverse-design routes for this hyperuniform network structure search include optimization schemes (e.g., simulated annealing) or machine learning methods.
Another important area of future work is the development of efficient methods to construct and characterize the two-phase media associated with 3D networks to better study the structure and properties of 3D-printed network metamaterials.

~
\begin{acknowledgments}
The authors would like to thank J. Raj for assistance in the generation of network structures, M. Skolnick for insightful discussions, and M. A. Porter for valuable feedback on the manuscript.  This work is supported by the collaborative NSF DMREF Grant No. CMMI-2323342 and NSF Grant No. DMS-2307297.
\end{acknowledgments}

\section*{Data Availability}
The code and data that support the findings of this article are openly available \cite{Maher_GH,Maher_data}.

%


\begin{thebibliography}{40}%
\makeatletter
\providecommand \@ifxundefined [1]{%
 \@ifx{#1\undefined}
}%
\providecommand \@ifnum [1]{%
 \ifnum #1\expandafter \@firstoftwo
 \else \expandafter \@secondoftwo
 \fi
}%
\providecommand \@ifx [1]{%
 \ifx #1\expandafter \@firstoftwo
 \else \expandafter \@secondoftwo
 \fi
}%
\providecommand \natexlab [1]{#1}%
\providecommand \enquote  [1]{``#1''}%
\providecommand \bibnamefont  [1]{#1}%
\providecommand \bibfnamefont [1]{#1}%
\providecommand \citenamefont [1]{#1}%
\providecommand \href@noop [0]{\@secondoftwo}%
\providecommand \href [0]{\begingroup \@sanitize@url \@href}%
\providecommand \@href[1]{\@@startlink{#1}\@@href}%
\providecommand \@@href[1]{\endgroup#1\@@endlink}%
\providecommand \@sanitize@url [0]{\catcode `\\12\catcode `\$12\catcode `\&12\catcode `\#12\catcode `\^12\catcode `\_12\catcode `\%12\relax}%
\providecommand \@@startlink[1]{}%
\providecommand \@@endlink[0]{}%
\providecommand \url  [0]{\begingroup\@sanitize@url \@url }%
\providecommand \@url [1]{\endgroup\@href {#1}{\urlprefix }}%
\providecommand \urlprefix  [0]{URL }%
\providecommand \Eprint [0]{\href }%
\providecommand \doibase [0]{https://doi.org/}%
\providecommand \selectlanguage [0]{\@gobble}%
\providecommand \bibinfo  [0]{\@secondoftwo}%
\providecommand \bibfield  [0]{\@secondoftwo}%
\providecommand \translation [1]{[#1]}%
\providecommand \BibitemOpen [0]{}%
\providecommand \bibitemStop [0]{}%
\providecommand \bibitemNoStop [0]{.\EOS\space}%
\providecommand \EOS [0]{\spacefactor3000\relax}%
\providecommand \BibitemShut  [1]{\csname bibitem#1\endcsname}%
\let\auto@bib@innerbib\@empty
\bibitem [{\citenamefont {Smith}\ \emph {et~al.}(2004)\citenamefont {Smith}, \citenamefont {Pendry},\ and\ \citenamefont {Wiltshire}}]{smith_metamaterials_2004}%
  \BibitemOpen
  \bibfield  {author} {\bibinfo {author} {\bibfnamefont {D.~R.}\ \bibnamefont {Smith}}, \bibinfo {author} {\bibfnamefont {J.~B.}\ \bibnamefont {Pendry}},\ and\ \bibinfo {author} {\bibfnamefont {M.~C.~K.}\ \bibnamefont {Wiltshire}},\ }\bibfield  {title} {\bibinfo {title} {Metamaterials and negative refractive index},\ }\href {https://doi.org/10.1126/science.1096796} {\bibfield  {journal} {\bibinfo  {journal} {Science}\ }\textbf {\bibinfo {volume} {305}},\ \bibinfo {pages} {788} (\bibinfo {year} {2004})}\BibitemShut {NoStop}%
\bibitem [{\citenamefont {Dubey}\ \emph {et~al.}(2024)\citenamefont {Dubey}, \citenamefont {Mirhakimi},\ and\ \citenamefont {Elbestawi}}]{dubey_negative_2024}%
  \BibitemOpen
  \bibfield  {author} {\bibinfo {author} {\bibfnamefont {D.}~\bibnamefont {Dubey}}, \bibinfo {author} {\bibfnamefont {A.~S.}\ \bibnamefont {Mirhakimi}},\ and\ \bibinfo {author} {\bibfnamefont {M.~A.}\ \bibnamefont {Elbestawi}},\ }\bibfield  {title} {\bibinfo {title} {Negative thermal expansion metamaterials: a review of design, fabrication, and applications},\ }\href {https://doi.org/10.3390/jmmp8010040} {\bibfield  {journal} {\bibinfo  {journal} {Journal of Manufacturing and Materials Processing}\ }\textbf {\bibinfo {volume} {8}},\ \bibinfo {pages} {40} (\bibinfo {year} {2024})}\BibitemShut {NoStop}%
\bibitem [{\citenamefont {Sniechowski}\ \emph {et~al.}(2015)\citenamefont {Sniechowski}, \citenamefont {Kaminski},\ and\ \citenamefont {Wronski}}]{sniechowski_heterogeneous_2015}%
  \BibitemOpen
  \bibfield  {author} {\bibinfo {author} {\bibfnamefont {M.}~\bibnamefont {Sniechowski}}, \bibinfo {author} {\bibfnamefont {J.}~\bibnamefont {Kaminski}},\ and\ \bibinfo {author} {\bibfnamefont {S.}~\bibnamefont {Wronski}},\ }\bibfield  {title} {\bibinfo {title} {Heterogeneous materials based on aperiodic structures for bone tissue substitutes},\ }\href@noop {} {\bibfield  {journal} {\bibinfo  {journal} {VI International Conference on Computational Bioengineering ICCB}\ } (\bibinfo {year} {2015})}\BibitemShut {NoStop}%
\bibitem [{\citenamefont {Mohammed}\ and\ \citenamefont {Gibson}(2018)}]{mohammed_design_2018}%
  \BibitemOpen
  \bibfield  {author} {\bibinfo {author} {\bibfnamefont {M.~I.}\ \bibnamefont {Mohammed}}\ and\ \bibinfo {author} {\bibfnamefont {I.}~\bibnamefont {Gibson}},\ }\bibfield  {title} {\bibinfo {title} {Design of three-dimensional, triply periodic unit cell scaffold structures for additive manufacturing},\ }\href {https://doi.org/10.1115/1.4040164} {\bibfield  {journal} {\bibinfo  {journal} {Journal of Mechanical Design}\ }\textbf {\bibinfo {volume} {140}},\ \bibinfo {pages} {071701} (\bibinfo {year} {2018})}\BibitemShut {NoStop}%
\bibitem [{\citenamefont {Ramírez-Torres}\ \emph {et~al.}(2018)\citenamefont {Ramírez-Torres}, \citenamefont {Penta}, \citenamefont {Rodríguez-Ramos}, \citenamefont {Merodio}, \citenamefont {Sabina}, \citenamefont {Bravo-Castillero}, \citenamefont {Guinovart-Díaz}, \citenamefont {Preziosi},\ and\ \citenamefont {Grillo}}]{ramirez-torres_three_2018}%
  \BibitemOpen
  \bibfield  {author} {\bibinfo {author} {\bibfnamefont {A.}~\bibnamefont {Ramírez-Torres}}, \bibinfo {author} {\bibfnamefont {R.}~\bibnamefont {Penta}}, \bibinfo {author} {\bibfnamefont {R.}~\bibnamefont {Rodríguez-Ramos}}, \bibinfo {author} {\bibfnamefont {J.}~\bibnamefont {Merodio}}, \bibinfo {author} {\bibfnamefont {F.~J.}\ \bibnamefont {Sabina}}, \bibinfo {author} {\bibfnamefont {J.}~\bibnamefont {Bravo-Castillero}}, \bibinfo {author} {\bibfnamefont {R.}~\bibnamefont {Guinovart-Díaz}}, \bibinfo {author} {\bibfnamefont {L.}~\bibnamefont {Preziosi}},\ and\ \bibinfo {author} {\bibfnamefont {A.}~\bibnamefont {Grillo}},\ }\bibfield  {title} {\bibinfo {title} {Three scales asymptotic homogenization and its application to layered hierarchical hard tissues},\ }\href {https://doi.org/10.1016/j.ijsolstr.2017.09.035} {\bibfield  {journal} {\bibinfo  {journal} {International Journal of Solids and Structures}\ }\textbf {\bibinfo {volume} {130-131}},\ \bibinfo {pages} {190} (\bibinfo {year} {2018})}\BibitemShut
  {NoStop}%
\bibitem [{\citenamefont {Wit}\ \emph {et~al.}(2019)\citenamefont {Wit}, \citenamefont {Wronski},\ and\ \citenamefont {Tarasiuk}}]{wit_simulation_2019}%
  \BibitemOpen
  \bibfield  {author} {\bibinfo {author} {\bibfnamefont {A.}~\bibnamefont {Wit}}, \bibinfo {author} {\bibfnamefont {S.}~\bibnamefont {Wronski}},\ and\ \bibinfo {author} {\bibfnamefont {J.}~\bibnamefont {Tarasiuk}},\ }\bibfield  {title} {\bibinfo {title} {Simulation and optimization of porous bone-like microstructures with specific mechanical properties},\ }\href {https://doi.org/10.14311/APP.2019.25.0089} {\bibfield  {journal} {\bibinfo  {journal} {Acta Polytechnica CTU Proceedings}\ }\textbf {\bibinfo {volume} {25}},\ \bibinfo {pages} {89} (\bibinfo {year} {2019})}\BibitemShut {NoStop}%
\bibitem [{\citenamefont {Zachary}\ and\ \citenamefont {Torquato}(2009)}]{zachary_hyperuniformity_2009}%
  \BibitemOpen
  \bibfield  {author} {\bibinfo {author} {\bibfnamefont {C.~E.}\ \bibnamefont {Zachary}}\ and\ \bibinfo {author} {\bibfnamefont {S.}~\bibnamefont {Torquato}},\ }\bibfield  {title} {\bibinfo {title} {Hyperuniformity in point patterns and two-phase random heterogeneous media},\ }\href {https://doi.org/10.1088/1742-5468/2009/12/P12015} {\bibfield  {journal} {\bibinfo  {journal} {Journal of Statistical Mechanics: Theory and Experiment}\ }\textbf {\bibinfo {volume} {2009}},\ \bibinfo {pages} {P12015} (\bibinfo {year} {2009})}\BibitemShut {NoStop}%
\bibitem [{\citenamefont {Torquato}(2018)}]{torquato_hyperuniform_2018}%
  \BibitemOpen
  \bibfield  {author} {\bibinfo {author} {\bibfnamefont {S.}~\bibnamefont {Torquato}},\ }\bibfield  {title} {\bibinfo {title} {Hyperuniform states of matter},\ }\href {https://doi.org/10.1016/j.physrep.2018.03.001} {\bibfield  {journal} {\bibinfo  {journal} {Physics Reports}\ }\bibinfo {series} {Hyperuniform {States} of {Matter}},\ \textbf {\bibinfo {volume} {745}},\ \bibinfo {pages} {1} (\bibinfo {year} {2018})}\BibitemShut {NoStop}%
\bibitem [{\citenamefont {Florescu}\ \emph {et~al.}(2009)\citenamefont {Florescu}, \citenamefont {Torquato},\ and\ \citenamefont {Steinhardt}}]{florescu_designer_2009}%
  \BibitemOpen
  \bibfield  {author} {\bibinfo {author} {\bibfnamefont {M.}~\bibnamefont {Florescu}}, \bibinfo {author} {\bibfnamefont {S.}~\bibnamefont {Torquato}},\ and\ \bibinfo {author} {\bibfnamefont {P.~J.}\ \bibnamefont {Steinhardt}},\ }\bibfield  {title} {\bibinfo {title} {Designer disordered materials with large, complete photonic band gaps},\ }\href {https://doi.org/10.1073/pnas.0907744106} {\bibfield  {journal} {\bibinfo  {journal} {Proceedings of the National Academy of Sciences}\ }\textbf {\bibinfo {volume} {106}},\ \bibinfo {pages} {20658} (\bibinfo {year} {2009})}\BibitemShut {NoStop}%
\bibitem [{\citenamefont {Froufe-Pérez}\ \emph {et~al.}(2016)\citenamefont {Froufe-Pérez}, \citenamefont {Engel}, \citenamefont {Damasceno}, \citenamefont {Muller}, \citenamefont {Haberko}, \citenamefont {Glotzer},\ and\ \citenamefont {Scheffold}}]{froufe-perez_role_2016}%
  \BibitemOpen
  \bibfield  {author} {\bibinfo {author} {\bibfnamefont {L.~S.}\ \bibnamefont {Froufe-Pérez}}, \bibinfo {author} {\bibfnamefont {M.}~\bibnamefont {Engel}}, \bibinfo {author} {\bibfnamefont {P.~F.}\ \bibnamefont {Damasceno}}, \bibinfo {author} {\bibfnamefont {N.}~\bibnamefont {Muller}}, \bibinfo {author} {\bibfnamefont {J.}~\bibnamefont {Haberko}}, \bibinfo {author} {\bibfnamefont {S.~C.}\ \bibnamefont {Glotzer}},\ and\ \bibinfo {author} {\bibfnamefont {F.}~\bibnamefont {Scheffold}},\ }\bibfield  {title} {\bibinfo {title} {Role of short-range order and hyperuniformity in the formation of band gaps in disordered photonic materials},\ }\href {https://doi.org/10.1103/PhysRevLett.117.053902} {\bibfield  {journal} {\bibinfo  {journal} {Physical Review Letters}\ }\textbf {\bibinfo {volume} {117}},\ \bibinfo {pages} {053902} (\bibinfo {year} {2016})}\BibitemShut {NoStop}%
\bibitem [{\citenamefont {Gkantzounis}\ \emph {et~al.}(2017)\citenamefont {Gkantzounis}, \citenamefont {Amoah},\ and\ \citenamefont {Florescu}}]{gkantzounis_hyperuniform_2017}%
  \BibitemOpen
  \bibfield  {author} {\bibinfo {author} {\bibfnamefont {G.}~\bibnamefont {Gkantzounis}}, \bibinfo {author} {\bibfnamefont {T.}~\bibnamefont {Amoah}},\ and\ \bibinfo {author} {\bibfnamefont {M.}~\bibnamefont {Florescu}},\ }\bibfield  {title} {\bibinfo {title} {Hyperuniform disordered phononic structures},\ }\href {https://doi.org/10.1103/PhysRevB.95.094120} {\bibfield  {journal} {\bibinfo  {journal} {Physical Review B}\ }\textbf {\bibinfo {volume} {95}},\ \bibinfo {pages} {094120} (\bibinfo {year} {2017})}\BibitemShut {NoStop}%
\bibitem [{\citenamefont {Siedentop}\ \emph {et~al.}(2024)\citenamefont {Siedentop}, \citenamefont {Lui}, \citenamefont {Maret}, \citenamefont {Chaikin}, \citenamefont {Steinhardt}, \citenamefont {Torquato}, \citenamefont {Keim},\ and\ \citenamefont {Florescu}}]{siedentop_stealthy_2024}%
  \BibitemOpen
  \bibfield  {author} {\bibinfo {author} {\bibfnamefont {L.}~\bibnamefont {Siedentop}}, \bibinfo {author} {\bibfnamefont {G.}~\bibnamefont {Lui}}, \bibinfo {author} {\bibfnamefont {G.}~\bibnamefont {Maret}}, \bibinfo {author} {\bibfnamefont {P.~M.}\ \bibnamefont {Chaikin}}, \bibinfo {author} {\bibfnamefont {P.~J.}\ \bibnamefont {Steinhardt}}, \bibinfo {author} {\bibfnamefont {S.}~\bibnamefont {Torquato}}, \bibinfo {author} {\bibfnamefont {P.}~\bibnamefont {Keim}},\ and\ \bibinfo {author} {\bibfnamefont {M.}~\bibnamefont {Florescu}},\ } \bibfield  {title} {\bibinfo {title} {Stealthy and hyperuniform isotropic photonic band gap structure in {3D}}},\ \href {https://doi.org/10.1093/pnasnexus/pgae383} {\bibfield  {journal} {\bibinfo  {journal} {PNAS Nexus}\ }\textbf {\bibinfo {volume} {3}},\ \bibinfo {pages} {pgae383} (\bibinfo {year} {2024})}\BibitemShut {NoStop}%
\bibitem [{\citenamefont {Torquato}(2021)}]{torquato_diffusion_2021}%
  \BibitemOpen
  \bibfield  {author} {\bibinfo {author} {\bibfnamefont {S.}~\bibnamefont {Torquato}},\ }\bibfield  {title} {\bibinfo {title} {Diffusion spreadability as a probe of the microstructure of complex media across length scales},\ }\href {https://doi.org/10.1103/PhysRevE.104.054102} {\bibfield  {journal} {\bibinfo  {journal} {Physical Review E}\ }\textbf {\bibinfo {volume} {104}},\ \bibinfo {pages} {054102} (\bibinfo {year} {2021})}\BibitemShut {NoStop}%
\bibitem [{\citenamefont {Torquato}\ and\ \citenamefont {Chen}(2018)}]{torquato_multifunctional_2018}%
  \BibitemOpen
  \bibfield  {author} {\bibinfo {author} {\bibfnamefont {S.}~\bibnamefont {Torquato}}\ and\ \bibinfo {author} {\bibfnamefont {D.}~\bibnamefont {Chen}},\ }\bibfield  {title} {\bibinfo {title} {Multifunctional hyperuniform cellular networks: optimality, anisotropy and disorder},\ }\href {https://doi.org/10.1088/2399-7532/aaca91} {\bibfield  {journal} {\bibinfo  {journal} {Multifunctional Materials}\ }\textbf {\bibinfo {volume} {1}},\ \bibinfo {pages} {015001} (\bibinfo {year} {2018})}\BibitemShut {NoStop}%
\bibitem [{\citenamefont {Kim}\ and\ \citenamefont {Torquato}(2019)}]{kim_new_2019}%
  \BibitemOpen
  \bibfield  {author} {\bibinfo {author} {\bibfnamefont {J.}~\bibnamefont {Kim}}\ and\ \bibinfo {author} {\bibfnamefont {S.}~\bibnamefont {Torquato}},\ }\bibfield  {title} {\bibinfo {title} {New tessellation-based procedure to design perfectly hyperuniform disordered dispersions for materials discovery},\ }\href {https://doi.org/10.1016/j.actamat.2019.01.026} {\bibfield  {journal} {\bibinfo  {journal} {Acta Materialia}\ }\textbf {\bibinfo {volume} {168}},\ \bibinfo {pages} {143} (\bibinfo {year} {2019})}\BibitemShut {NoStop}%
\bibitem [{\citenamefont {McGregor}\ \emph {et~al.}(2024)\citenamefont {McGregor}, \citenamefont {Patel}, \citenamefont {Zhang}, \citenamefont {Yu}, \citenamefont {Vlasea},\ and\ \citenamefont {McLachlin}}]{mcgregor_manufacturability_2024}%
  \BibitemOpen
  \bibfield  {author} {\bibinfo {author} {\bibfnamefont {M.}~\bibnamefont {McGregor}}, \bibinfo {author} {\bibfnamefont {S.}~\bibnamefont {Patel}}, \bibinfo {author} {\bibfnamefont {K.}~\bibnamefont {Zhang}}, \bibinfo {author} {\bibfnamefont {A.}~\bibnamefont {Yu}}, \bibinfo {author} {\bibfnamefont {M.}~\bibnamefont {Vlasea}},\ and\ \bibinfo {author} {\bibfnamefont {S.}~\bibnamefont {McLachlin}},\ }\bibfield  {title} {\bibinfo {title} {A manufacturability evaluation of complex architectures by laser powder bed fusion additive manufacturing},\ }\href {https://doi.org/10.1115/1.4065315} {\bibfield  {journal} {\bibinfo  {journal} {Journal of Manufacturing Science and Engineering}\ }\textbf {\bibinfo {volume} {146}},\ \bibinfo {pages} {061007} (\bibinfo {year} {2024})}\BibitemShut {NoStop}%
\bibitem [{\citenamefont {Obrero}\ \emph {et~al.}(2024)\citenamefont {Obrero}, \citenamefont {Tirfe}, \citenamefont {Lee}, \citenamefont {Saptarshi}, \citenamefont {Rock}, \citenamefont {Daniels},\ and\ \citenamefont {Newhall}}]{obrero_electrical_2024}%
  \BibitemOpen
  \bibfield  {author} {\bibinfo {author} {\bibfnamefont {C.}~\bibnamefont {Obrero}}, \bibinfo {author} {\bibfnamefont {M.}~\bibnamefont {Tirfe}}, \bibinfo {author} {\bibfnamefont {C.}~\bibnamefont {Lee}}, \bibinfo {author} {\bibfnamefont {S.}~\bibnamefont {Saptarshi}}, \bibinfo {author} {\bibfnamefont {C.}~\bibnamefont {Rock}}, \bibinfo {author} {\bibfnamefont {K.~E.}\ \bibnamefont {Daniels}},\ and\ \bibinfo {author} {\bibfnamefont {K.~A.}\ \bibnamefont {Newhall}},\ }\href {https://doi.org/10.48550/arXiv.2410.11525} {\bibinfo {title} {Electrical transport in tunably-disordered metamaterials}} (\bibinfo {year} {2024}),\ \bibinfo {note} {arXiv:2410.11525}\BibitemShut {NoStop}%
\bibitem [{\citenamefont {Torquato}\ and\ \citenamefont {Stillinger}(2003)}]{torquato_local_2003}%
  \BibitemOpen
  \bibfield  {author} {\bibinfo {author} {\bibfnamefont {S.}~\bibnamefont {Torquato}}\ and\ \bibinfo {author} {\bibfnamefont {F.~H.}\ \bibnamefont {Stillinger}},\ }\bibfield  {title} {\bibinfo {title} {Local density fluctuations, hyperuniformity, and order metrics},\ }\href {https://doi.org/10.1103/PhysRevE.68.041113} {\bibfield  {journal} {\bibinfo  {journal} {Physical Review E}\ }\textbf {\bibinfo {volume} {68}},\ \bibinfo {pages} {041113} (\bibinfo {year} {2003})}\BibitemShut {NoStop}%
\bibitem [{\citenamefont {Torquato}\ \emph {et~al.}(2015)\citenamefont {Torquato}, \citenamefont {Zhang},\ and\ \citenamefont {Stillinger}}]{torquato_ensemble_2015}%
  \BibitemOpen
  \bibfield  {author} {\bibinfo {author} {\bibfnamefont {S.}~\bibnamefont {Torquato}}, \bibinfo {author} {\bibfnamefont {G.}~\bibnamefont {Zhang}},\ and\ \bibinfo {author} {\bibfnamefont {F.~H.}\ \bibnamefont {Stillinger}},\ }\bibfield  {title} {\bibinfo {title} {Ensemble theory for stealthy hyperuniform disordered ground states},\ }\href {https://doi.org/10.1103/PhysRevX.5.021020} {\bibfield  {journal} {\bibinfo  {journal} {Physical Review X}\ }\textbf {\bibinfo {volume} {5}},\ \bibinfo {pages} {021020} (\bibinfo {year} {2015})}\BibitemShut {NoStop}%
\bibitem [{\citenamefont {Chen}\ \emph {et~al.}(2018)\citenamefont {Chen}, \citenamefont {Lomba},\ and\ \citenamefont {Torquato}}]{chen_binary_2018}%
  \BibitemOpen
  \bibfield  {author} {\bibinfo {author} {\bibfnamefont {D.}~\bibnamefont {Chen}}, \bibinfo {author} {\bibfnamefont {E.}~\bibnamefont {Lomba}},\ and\ \bibinfo {author} {\bibfnamefont {S.}~\bibnamefont {Torquato}},\ }\bibfield  {title} {\bibinfo {title} {Binary mixtures of charged colloids: a potential route to synthesize disordered hyperuniform materials},\ }\href {https://doi.org/10.1039/C8CP02616E} {\bibfield  {journal} {\bibinfo  {journal} {Physical Chemistry Chemical Physics}\ }\textbf {\bibinfo {volume} {20}},\ \bibinfo {pages} {17557} (\bibinfo {year} {2018})}\BibitemShut {NoStop}%
\bibitem [{\citenamefont {Maher}\ \emph {et~al.}(2022)\citenamefont {Maher}, \citenamefont {Stillinger},\ and\ \citenamefont {Torquato}}]{maher_characterization_2022}%
  \BibitemOpen
  \bibfield  {author} {\bibinfo {author} {\bibfnamefont {C.~E.}\ \bibnamefont {Maher}}, \bibinfo {author} {\bibfnamefont {F.~H.}\ \bibnamefont {Stillinger}},\ and\ \bibinfo {author} {\bibfnamefont {S.}~\bibnamefont {Torquato}},\ }\bibfield  {title} {\bibinfo {title} {Characterization of void space, large-scale structure, and transport properties of maximally random jammed packings of superballs},\ }\href {https://doi.org/10.1103/PhysRevMaterials.6.025603} {\bibfield  {journal} {\bibinfo  {journal} {Physical Review Materials}\ }\textbf {\bibinfo {volume} {6}},\ \bibinfo {pages} {025603} (\bibinfo {year} {2022})}\BibitemShut {NoStop}%
\bibitem [{\citenamefont {Torquato}(2016)}]{torquato_disordered_2016}%
  \BibitemOpen
  \bibfield  {author} {\bibinfo {author} {\bibfnamefont {S.}~\bibnamefont {Torquato}},\ }\bibfield  {title} {\bibinfo {title} {Disordered hyperuniform heterogeneous materials},\ }\href {https://doi.org/10.1088/0953-8984/28/41/414012} {\bibfield  {journal} {\bibinfo  {journal} {Journal of Physics: Condensed Matter}\ }\textbf {\bibinfo {volume} {28}},\ \bibinfo {pages} {414012} (\bibinfo {year} {2016})}\BibitemShut {NoStop}%
\bibitem [{\citenamefont {Man}\ \emph {et~al.}(2013)\citenamefont {Man}, \citenamefont {Florescu}, \citenamefont {Williamson}, \citenamefont {He}, \citenamefont {Hashemizad}, \citenamefont {Leung}, \citenamefont {Liner}, \citenamefont {Torquato}, \citenamefont {Chaikin},\ and\ \citenamefont {Steinhardt}}]{man_isotropic_2013}%
  \BibitemOpen
  \bibfield  {author} {\bibinfo {author} {\bibfnamefont {W.}~\bibnamefont {Man}}, \bibinfo {author} {\bibfnamefont {M.}~\bibnamefont {Florescu}}, \bibinfo {author} {\bibfnamefont {E.~P.}\ \bibnamefont {Williamson}}, \bibinfo {author} {\bibfnamefont {Y.}~\bibnamefont {He}}, \bibinfo {author} {\bibfnamefont {S.~R.}\ \bibnamefont {Hashemizad}}, \bibinfo {author} {\bibfnamefont {B.~Y.~C.}\ \bibnamefont {Leung}}, \bibinfo {author} {\bibfnamefont {D.~R.}\ \bibnamefont {Liner}}, \bibinfo {author} {\bibfnamefont {S.}~\bibnamefont {Torquato}}, \bibinfo {author} {\bibfnamefont {P.~M.}\ \bibnamefont {Chaikin}},\ and\ \bibinfo {author} {\bibfnamefont {P.~J.}\ \bibnamefont {Steinhardt}},\ }\bibfield  {title} {\bibinfo {title} {Isotropic band gaps and freeform waveguides observed in hyperuniform disordered photonic solids},\ }\href {https://doi.org/10.1073/pnas.1307879110} {\bibfield  {journal} {\bibinfo  {journal} {Proceedings of the National Academy of Sciences}\ }\textbf {\bibinfo {volume} {110}},\ \bibinfo {pages}
  {15886} (\bibinfo {year} {2013})}\BibitemShut {NoStop}%
\bibitem [{\citenamefont {Raj}\ \emph{et al.}(2025)\citenamefont{Raj},\ and\ \citenamefont{\emph{et al.}}}]{raj_upcoming_2025}%
  \BibitemOpen
  \bibfield  {author} {\bibinfo {author} {\bibfnamefont {J.}~\bibnamefont {Raj}}, \bibinfo {author} {\bibfnamefont {\emph{et.}}~\bibnamefont {\emph{al}}},\ }\href@noop {} \bibfield  {title} {\bibinfo {title} {Local Geometric and Transport Properties of Spatially Embedded Networks Generated by Hyperuniform Point Patterns},} {\  (\bibinfo {year} {forthcoming})}\BibitemShut {NoStop}%
\bibitem [{\citenamefont {Chen}\ \emph {et~al.}(2021{\natexlab{a}})\citenamefont {Chen}, \citenamefont {Zheng}, \citenamefont {Liu}, \citenamefont {Zhang}, \citenamefont {Chen}, \citenamefont {Jiao},\ and\ \citenamefont {Zhuang}}]{chen_stonewales_2021}%
  \BibitemOpen
  \bibfield  {author} {\bibinfo {author} {\bibfnamefont {D.}~\bibnamefont {Chen}}, \bibinfo {author} {\bibfnamefont {Y.}~\bibnamefont {Zheng}}, \bibinfo {author} {\bibfnamefont {L.}~\bibnamefont {Liu}}, \bibinfo {author} {\bibfnamefont {G.}~\bibnamefont {Zhang}}, \bibinfo {author} {\bibfnamefont {M.}~\bibnamefont {Chen}}, \bibinfo {author} {\bibfnamefont {Y.}~\bibnamefont {Jiao}},\ and\ \bibinfo {author} {\bibfnamefont {H.}~\bibnamefont {Zhuang}},\ }\bibfield  {title} {{\bibinfo {title} {Stone–{Wales} defects preserve hyperuniformity in amorphous two-dimensional networks}},\ }\href {https://doi.org/10.1073/pnas.2016862118}{\bibfield  {journal} {\bibinfo  {journal} {Proceedings of the National Academy of Sciences}\ }\textbf {\bibinfo {volume} {118}},\  (\bibinfo {year} {2021}{\natexlab{a}})}\BibitemShut {NoStop}%
\bibitem [{\citenamefont {Chen}\ \emph {et~al.}(2021{\natexlab{b}})\citenamefont {Chen}, \citenamefont {Zheng},\ and\ \citenamefont {Jiao}}]{chen_topological_2021}%
  \BibitemOpen
  \bibfield  {author} {\bibinfo {author} {\bibfnamefont {D.}~\bibnamefont {Chen}}, \bibinfo {author} {\bibfnamefont {Y.}~\bibnamefont {Zheng}},\ and\ \bibinfo {author} {\bibfnamefont {Y.}~\bibnamefont {Jiao}},\ }\bibfield  {title} {\bibinfo {title} {Topological defects, inherent structures, and hyperuniformity},\ }\href {https://doi.org/10.1103/PhysRevB.104.174101} {\bibfield  {journal} {\bibinfo  {journal} {Physical Review B}\ }\textbf {\bibinfo {volume} {104}},\ \bibinfo {pages} {174101} (\bibinfo {year} {2021}{\natexlab{b}})}\BibitemShut {NoStop}%
\bibitem [{\citenamefont {Salvalaglio}\ \emph {et~al.}(2024)\citenamefont {Salvalaglio}, \citenamefont {Skinner}, \citenamefont {Dunkel},\ and\ \citenamefont {Voigt}}]{salvalaglio_persistent_2024a}%
  \BibitemOpen
  \bibfield  {author} {\bibinfo {author} {\bibfnamefont {M.}~\bibnamefont {Salvalaglio}}, \bibinfo {author} {\bibfnamefont {D.~J.}\ \bibnamefont {Skinner}}, \bibinfo {author} {\bibfnamefont {J.}~\bibnamefont {Dunkel}},\ and\ \bibinfo {author} {\bibfnamefont {A.}~\bibnamefont {Voigt}},\ }\bibfield  {title} {\bibinfo {title} {Persistent homology and topological statistics of hyperuniform point clouds},\ }\href {https://doi.org/10.1103/PhysRevResearch.6.023107} {\bibfield  {journal} {\bibinfo  {journal} {Physical Review Research}\ }\textbf {\bibinfo {volume} {6}},\ \bibinfo {pages} {023107} (\bibinfo {year} {2024})}\BibitemShut {NoStop}%
\bibitem [{\citenamefont {Newby}\ \emph {et~al.}(2024)\citenamefont {Newby}, \citenamefont {Shi}, \citenamefont {Jiao}, \citenamefont {Albert},\ and\ \citenamefont {Torquato}}]{newby_structural_2024}%
  \BibitemOpen
  \bibfield  {author} {\bibinfo {author} {\bibfnamefont {E.}~\bibnamefont {Newby}}, \bibinfo {author} {\bibfnamefont {W.}~\bibnamefont {Shi}}, \bibinfo {author} {\bibfnamefont {Y.}~\bibnamefont {Jiao}}, \bibinfo {author} {\bibfnamefont {R.}~\bibnamefont {Albert}},\ and\ \bibinfo {author} {\bibfnamefont {S.}~\bibnamefont {Torquato}},\ }\href {https://doi.org/10.1103/PhysRevE.111.034123}  {\bibinfo {title} {Structural Properties of Hyperuniform Networks}} {\bibfield  {journal} {\bibinfo  {journal} {Physical Review E}\ }\textbf {\bibinfo {volume} {111}},\ \bibinfo {pages} {034123} (\bibinfo {year} {2025})}\BibitemShut {NoStop}%
\bibitem [{\citenamefont {Newby}\ \emph {et~al.}(2025)\citenamefont {Newby}, \citenamefont {Shi}, \citenamefont {Jiao}, \citenamefont {Torquato},\ and\ \citenamefont {Albert}}]{newby_point_2025}%
  \BibitemOpen
  \bibfield  {author} {\bibinfo {author} {\bibfnamefont {E.}~\bibnamefont {Newby}}, \bibinfo {author} {\bibfnamefont {W.}~\bibnamefont {Shi}}, \bibinfo {author} {\bibfnamefont {Y.}~\bibnamefont {Jiao}}, \bibinfo {author} {\bibfnamefont {S.}~\bibnamefont {Torquato}},\ and\ \bibinfo {author} {\bibfnamefont {R.}~\bibnamefont {Albert}},\ }\href {https://doi.org/10.48550/arXiv.2504.01015} {\bibinfo {title} {From point patterns to networks: to what extent does the {Delaunay} triangulation reproduce key spatial and density information?}} (\bibinfo {year} {2025}),\ \bibinfo {note} {arXiv:2504.01015 [cond-mat]}\BibitemShut {NoStop}%
\bibitem [{\citenamefont {Klatt}\ \emph {et~al.}(2020)\citenamefont {Klatt}, \citenamefont {Kim},\ and\ \citenamefont {Torquato}}]{klatt_cloaking_2020}%
  \BibitemOpen
  \bibfield  {author} {\bibinfo {author} {\bibfnamefont {M.~A.}\ \bibnamefont {Klatt}}, \bibinfo {author} {\bibfnamefont {J.}~\bibnamefont {Kim}},\ and\ \bibinfo {author} {\bibfnamefont {S.}~\bibnamefont {Torquato}},\ }\bibfield  {title} {\bibinfo {title} {Cloaking the underlying long-range order of randomly perturbed lattices},\ }\href {https://doi.org/10.1103/PhysRevE.101.032118} {\bibfield  {journal} {\bibinfo  {journal} {Physical Review E}\ }\textbf {\bibinfo {volume} {101}},\ \bibinfo {pages} {032118} (\bibinfo {year} {2020})}\BibitemShut {NoStop}%
\bibitem [{\citenamefont {Torquato}\ \emph {et~al.}(2022)\citenamefont {Torquato}, \citenamefont {Skolnick},\ and\ \citenamefont {Kim}}]{torquato_local_2022}%
  \BibitemOpen
  \bibfield  {author} {\bibinfo {author} {\bibfnamefont {S.}~\bibnamefont {Torquato}}, \bibinfo {author} {\bibfnamefont {M.}~\bibnamefont {Skolnick}},\ and\ \bibinfo {author} {\bibfnamefont {J.}~\bibnamefont {Kim}},\ }\bibfield  {title} {\bibinfo {title} {Local order metrics for two-phase media across length scales},\ }\href {https://doi.org/10.1088/1751-8121/ac72d7} {\bibfield  {journal} {\bibinfo  {journal} {Journal of Physics A: Mathematical and Theoretical}\ }\textbf {\bibinfo {volume} {55}},\ \bibinfo {pages} {274003} (\bibinfo {year} {2022})}\BibitemShut {NoStop}%
\bibitem [{\citenamefont {Skolnick}\ and\ \citenamefont {Torquato}(2024)}]{skolnick_quantifying_2024}%
  \BibitemOpen
  \bibfield  {author} {\bibinfo {author} {\bibfnamefont {M.}~\bibnamefont {Skolnick}}\ and\ \bibinfo {author} {\bibfnamefont {S.}~\bibnamefont {Torquato}},\ }\bibfield  {title} {\bibinfo {title} {Quantifying phase mixing and separation behaviors across length and time scales},\ }\href {https://doi.org/10.1016/j.actamat.2024.119774} {\bibfield  {journal} {\bibinfo  {journal} {Acta Materialia}\ }\textbf {\bibinfo {volume} {268}},\ \bibinfo {pages} {119774} (\bibinfo {year} {2024})}\BibitemShut {NoStop}%
\bibitem [{\citenamefont {Maher}\ and\ \citenamefont {Torquato}(2024)}]{maher_local_2024}%
  \BibitemOpen
  \bibfield  {author} {\bibinfo {author} {\bibfnamefont {C.~E.}\ \bibnamefont {Maher}}\ and\ \bibinfo {author} {\bibfnamefont {S.}~\bibnamefont {Torquato}},\ }\bibfield  {title} {\bibinfo {title} {Local order metrics for many-particle systems across length scales},\ }\href {https://doi.org/10.1103/PhysRevResearch.6.033262} {\bibfield  {journal} {\bibinfo  {journal} {Physical Review Research}\ }\textbf {\bibinfo {volume} {6}},\ \bibinfo {pages} {033262} (\bibinfo {year} {2024})}\BibitemShut {NoStop}%
\bibitem [{\citenamefont {Hansen}\ and\ \citenamefont {McDonald}(1990)}]{hansen_theory_1990}%
  \BibitemOpen
  \bibfield  {author} {\bibinfo {author} {\bibfnamefont {J.-P.}\ \bibnamefont {Hansen}}\ and\ \bibinfo {author} {\bibfnamefont {I.~R.}\ \bibnamefont {McDonald}},\ }\href@noop {} {\emph {\bibinfo {title} {Theory of {Simple} {Liquids}}}}\ (\bibinfo  {publisher} {Elsevier},\ \bibinfo {year} {1990})\BibitemShut {NoStop}%
\bibitem [{\citenamefont {Torquato}(2002)}]{torquato_random_2002a}%
  \BibitemOpen
  \bibfield  {author} {\bibinfo {author} {\bibfnamefont {S.}~\bibnamefont {Torquato}},\ }\href {http://link.springer.com/10.1007/978-1-4757-6355-3} {\emph {\bibinfo {title} {Random {Heterogeneous} {Materials}}}},\ edited by\ \bibinfo {editor} {\bibfnamefont {S.~S.}\ \bibnamefont {Antman}}, \bibinfo {editor} {\bibfnamefont {L.}~\bibnamefont {Sirovich}}, \bibinfo {editor} {\bibfnamefont {J.~E.}\ \bibnamefont {Marsden}},\ and\ \bibinfo {editor} {\bibfnamefont {S.}~\bibnamefont {Wiggins}},\ \bibinfo {series} {Interdisciplinary {Applied} {Mathematics}}, Vol.~\bibinfo {volume} {16}\ (\bibinfo  {publisher} {Springer},\ \bibinfo {address} {New York, NY},\ \bibinfo {year} {2002})\BibitemShut {NoStop}%
\bibitem [{\citenamefont {Zachary}\ \emph {et~al.}(2011)\citenamefont {Zachary}, \citenamefont {Jiao},\ and\ \citenamefont {Torquato}}]{zachary_hyperuniformity_2011}%
  \BibitemOpen
  \bibfield  {author} {\bibinfo {author} {\bibfnamefont {C.~E.}\ \bibnamefont {Zachary}}, \bibinfo {author} {\bibfnamefont {Y.}~\bibnamefont {Jiao}},\ and\ \bibinfo {author} {\bibfnamefont {S.}~\bibnamefont {Torquato}},\ }\bibfield  {title} {\bibinfo {title} {Hyperuniformity, quasi-long-range correlations, and void-space constraints in maximally random jammed particle packings. {I}. {Polydisperse} spheres},\ }\href {https://doi.org/10.1103/PhysRevE.83.051308} {\bibfield  {journal} {\bibinfo  {journal} {Physical Review E}\ }\textbf {\bibinfo {volume} {83}},\ \bibinfo {pages} {051308} (\bibinfo {year} {2011})}\BibitemShut {NoStop}%
\bibitem [{\citenamefont {Zhang}\ \emph {et~al.}(2015)\citenamefont {Zhang}, \citenamefont {Stillinger},\ and\ \citenamefont {Torquato}}]{zhang_ground_2015}%
  \BibitemOpen
  \bibfield  {author} {\bibinfo {author} {\bibfnamefont {G.}~\bibnamefont {Zhang}}, \bibinfo {author} {\bibfnamefont {F.~H.}\ \bibnamefont {Stillinger}},\ and\ \bibinfo {author} {\bibfnamefont {S.}~\bibnamefont {Torquato}},\ }\bibfield  {title} {\bibinfo {title} {Ground states of stealthy hyperuniform potentials: {I}. {Entropically} favored configurations},\ }\href {https://doi.org/10.1103/PhysRevE.92.022119} {\bibfield  {journal} {\bibinfo  {journal} {Physical Review E}\ }\textbf {\bibinfo {volume} {92}},\ \bibinfo {pages} {022119} (\bibinfo {year} {2015})}\BibitemShut {NoStop}%
\bibitem [{\citenamefont {Gabriel}\ and\ \citenamefont {Sokal}(1969)}]{gabriel_new_1969}%
  \BibitemOpen
  \bibfield  {author} {\bibinfo {author} {\bibfnamefont {K.~R.}\ \bibnamefont {Gabriel}}\ and\ \bibinfo {author} {\bibfnamefont {R.~R.}\ \bibnamefont {Sokal}},\ }\bibfield  {title} {\bibinfo {title} {A new statistical approach to geographic variation analysis},\ }\href {https://doi.org/10.2307/2412323} {\bibfield  {journal} {\bibinfo  {journal} {Systematic Zoology}\ }\textbf {\bibinfo {volume} {18}},\ \bibinfo {pages} {259} (\bibinfo {year} {1969})}\BibitemShut {NoStop}%
\bibitem [{\citenamefont {Barthelemy}(2022)}]{barthelemy_spatial_2022}%
  \BibitemOpen
  \bibfield  {author} {\bibinfo {author} {\bibfnamefont {M.}~\bibnamefont {Barthelemy}},\ }\href {https://doi.org/10.1007/978-3-030-94106-2} {{\emph {\bibinfo {title} {Spatial {Networks}: {A} {Complete} {Introduction}: {From} {Graph} {Theory} and {Statistical} {Physics} to {Real}-{World} {Applications}}}}}\ (\bibinfo  {publisher} {Springer International Publishing},\ \bibinfo {address} {Cham},\ \bibinfo {year} {2022})\BibitemShut {NoStop}%
\bibitem [{\citenamefont {B\"ottcher}\ and\ \citenamefont {Porter}(2025)}]{bottcher_dynamical_2024}%
  \BibitemOpen
  \bibfield  {author} {\bibinfo {author} {\bibfnamefont {L.}~\bibnamefont {B\"ottcher}},\ and\ \bibinfo {author} {\bibfnamefont {M.~A.}~\bibnamefont {Porter}}}\ \href {https://doi.org/10.48550/arXiv.2401.00735} {\bibinfo {title} {Dynamical processes on metric networks}} (\bibinfo {year} {2024}),\ \bibinfo {note} {arXiv:2401.00735 [math.DS]}\BibitemShut {NoStop}%
\bibitem [{\citenamefont {Kim}\ and\ \citenamefont {Torquato}(2021)}]{kim_characterizing_2021}%
  \BibitemOpen
  \bibfield  {author} {\bibinfo {author} {\bibfnamefont {J.}~\bibnamefont {Kim}}\ and\ \bibinfo {author} {\bibfnamefont {S.}~\bibnamefont {Torquato}},\ }\bibfield  {title} {\bibinfo {title} {Characterizing the hyperuniformity of ordered and disordered two-phase media},\ }\href {https://doi.org/10.1103/PhysRevE.103.012123} {\bibfield  {journal} {\bibinfo  {journal} {Physical Review E}\ }\textbf {\bibinfo {volume} {103}},\ \bibinfo {pages} {012123} (\bibinfo {year} {2021})}\BibitemShut {NoStop}%
\bibitem [{\citenamefont {Liu}\ \emph {et~al.}(2024)\citenamefont {Liu}, \citenamefont {Chen}, \citenamefont {Tian}, \citenamefont {Xu},\ and\ \citenamefont {Jiao}}]{liu_universal_2024}%
  \BibitemOpen
  \bibfield  {author} {\bibinfo {author} {\bibfnamefont {Y.}~\bibnamefont {Liu}}, \bibinfo {author} {\bibfnamefont {D.}~\bibnamefont {Chen}}, \bibinfo {author} {\bibfnamefont {J.}~\bibnamefont {Tian}}, \bibinfo {author} {\bibfnamefont {W.}~\bibnamefont {Xu}},\ and\ \bibinfo {author} {\bibfnamefont {Y.}~\bibnamefont {Jiao}},\ }\bibfield  {title} {\bibinfo {title} {Universal {Hyperuniform} {Organization} in {Looped} {Leaf} {Vein} {Networks}},\ }\href {https://doi.org/10.1103/PhysRevLett.133.028401} {\bibfield  {journal} {\bibinfo  {journal} {Physical Review Letters}\ }\textbf {\bibinfo {volume} {133}},\ \bibinfo {pages} {028401} (\bibinfo {year} {2024})},\ \bibinfo {note} {publisher: American Physical Society}\BibitemShut {NoStop}%
\bibitem [{\citenamefont {Maher}()}]{Maher_GH}%
  \BibitemOpen
  \bibfield  {author} {\bibinfo {author} {\bibfnamefont {C. E.}~\bibnamefont
  {Maher}},\ }\href {https://github.com/DMREF-networks/config_stats} {\bibinfo
  {title} {https://github.com/DMREF-networks/Network\_Structure\_Characterization}}\BibitemShut {NoStop}%
\bibitem [{\citenamefont {Maher}\ and\ \citenamefont {Newhall}(2025)}]{Maher_data}%
  \BibitemOpen
  \bibfield  {author} {\bibinfo {author} {\bibfnamefont {C.~E.}\ \bibnamefont {Maher}}\ and\ \bibinfo {author} {\bibfnamefont {K.~A.}\ \bibnamefont {Newhall}},\ }\bibfield  {title} {\bibinfo {title} {Data from: {C}haracterizing the Hyperuniformity of Disordered Network Metamaterials}\ (\bibinfo {year} {Dryad, 2025}),\ }\href {https://doi.org/10.2307/2412323} {\bibinfo {note} {doi: 10.5061/dryad.tdz08kq9c}}\BibitemShut {NoStop}%
\end{thebibliography}
\end{document}